\documentclass[
a4paper,
aps,
pra,
showpacs,
twocolumn,
superscriptaddress]
{revtex4-2} 

\usepackage[utf8]{inputenc}  %Input what you want e.g., é, ł, a, ü
\usepackage[T1]{fontenc}     %Output what you want e.g., é, ł, a, ü
\usepackage[american]{babel}  %Do hyphenation according to american english
\usepackage[svgnames,dvipsnames]{xcolor}
\usepackage{CJKutf8}
\usepackage{utfsym}
\usepackage[colorlinks,
citecolor=DarkBlue,
linkcolor=DarkBlue,
linktocpage,
unicode]{hyperref}
\usepackage{amsmath,amssymb,amsthm,bm,mathtools,amsfonts,mathrsfs,bbm,dsfont} %Useful math packages
\newtheorem{theorem}{Theorem}

\newtheorem{result}[theorem]{Result}

\newcommand{\tr}{{\mathrm{tr}}}

\newcommand{\eins}{\mathbbm{1}}
\newcommand{\swap}{\mathbb{S}}
\renewcommand{\vr}{\ensuremath{\varrho}}
\renewcommand{\vec}[1]{\ensuremath{\boldsymbol{#1}}}
\usepackage{braket}
\usepackage[normalem]{ulem}
\usepackage{comment}
\newcommand{\ketbra}[2]{\ensuremath{| #1 \rangle\!\langle #2 |}}
\usepackage{url}

\begin{document}
\title{Entanglement in the symmetric subspace: mapping multipartite to bipartite states}

\author{Carlo Marconi}
\affiliation{Istituto Nazionale di Ottica - Consiglio Nazionale delle Ricerche (INO-CNR), Largo Enrico Fermi, 6, 50125 Firenze, Italy}

\author{Satoya Imai}
\affiliation{Institute of Systems and Information Engineering, University of Tsukuba, Tsukuba, Ibaraki 305-8573, Japan}
\affiliation{Center for Artificial Intelligence Research (C-AIR), University of Tsukuba, Tsukuba, Ibaraki 305-8577, Japan}
\affiliation{INO-CNR and LENS, Largo Enrico Fermi, 2, 50125 Firenze, Italy}

\date{\today}
\begin{abstract}
\noindent We propose a technique to investigate multipartite entanglement in the symmetric subspace. Our approach is to map an $N$-qubit symmetric state onto a bipartite symmetric state of higher local dimension. We show that this mapping preserves separability and allows to characterize the entanglement of the original multipartite state. In particular, we establish a connection between the border rank and the Schmidt rank, and derive lower bounds on entanglement measures. Finally, we reveal the existence of entangled symmetric subspaces, where all bipartite states are entangled.
\end{abstract}

\maketitle

%%%%%%%%%%%%%%%%%%%%%%%%%%%%%%%%%%%%%%%%%%%%%%%%%%%%%%%%%%%%%%%%
\section{Introduction}
%{\it Introduction.---}
Entanglement is a key resource in several quantum information processing tasks, ranging from  cryptography~\cite{ekert1991quantum,curty2004entanglement} to metrology~\cite{pezze2009entanglement} and measurement-based computation~\cite{raussendorf2001one}, among others. The rapid progress of experimental technology has enabled the generation of entangled states in various physical platforms (see Reviews~\cite{horodecki2009quantum,guhne2009entanglement,friis2019entanglement,erhard2020advances}). However, even for pure states, a complete characterization of multipartite entanglement remains challenging, for at least two reasons: i) a direct extension of the Schmidt decomposition in terms of multipartite orthogonal product states does not always exist~\cite{peres1995higher}; ii) multipartite maximally entangled states cannot be uniquely defined (see Review~\cite{plenio2005introduction}).

A strategy to address these issues is to focus on quantum systems whose state remains unaltered when any two particles are exchanged. Such states live in the \textit{symmetric subspace}, whose dimension is $N+1$ for $N$ qubits~\cite{harrow2013church,toth2009entanglement}. Considering the symmetric subspace can greatly simplify the computational complexity needed to characterize and quantify multipartite entanglement.

One example in this direction is given by the \textit{tensor rank}~\cite{eisert2001schmidt,chitambar2008tripartite}, a generalization of the Schmidt rank for multipartite systems. This is related to transformations between equivalence classes under stochastic local operations and classical communication (SLOCC)~\cite{dur2000three,lo2001concentrating}. In fact, although determining the tensor rank is known to be NP-hard~\cite{haastad1990tensor}, Ref.~\cite{chen2010tensor} showed that, for symmetric states, it can be efficiently estimated by the \textit{symmetric tensor rank} in the theory of homogeneous polynomials~\cite{comon2008symmetric}.

Another example is provided by the \textit{geometric measure of entanglement}, which quantifies how close a quantum state is to the set of separable states~\cite{shimony1995degree,wei2003geometric} (see Review~\cite{weinbrenner2025quantifying}). This measure has an operational interpretation, being connected to state discrimination via LOCC~\cite{hayashi2006bounds}. Even for pure states, calculating the geometric measure is generally a hard task~\cite{de2008tensor}, since it requires optimizing over the whole set of product states. In contrast, for symmetric states, Ref.~\cite{hubener2009geometric} has proven that it can be efficiently computed by restricting to the set of symmetric product states only.

%%%%%%%%%%%%%%%%%%%%%%%%%%%%%%%%%%%%%%%%%%%%%%%%%%%%%%%%%%%%%%%%
\begin{figure}[t]
    \centering
    \includegraphics[width=1.0\linewidth]{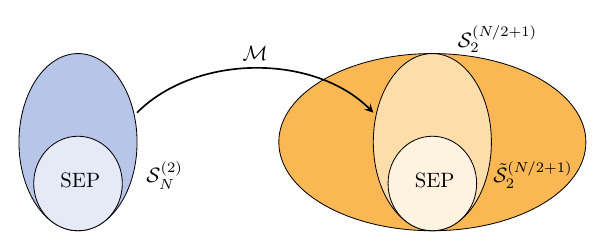}
    \caption{Pictorial representation of the mapping $\mathcal{M}$ presented in Result~\ref{ob:mapping}. This paper investigates the entanglement of multipartite states $\ket{\Psi} \in \mathcal{S}_{N}^{(2)}$ by focusing on their mapped bipartite states $\mathcal{M}(\ket{\Psi}) \in \tilde{\mathcal{S}}_{2}^{(N/2 +1)}$. Here SEP denotes the set of separable states. Note that, while all separable states in $\mathcal{S}_{N}^{(2)}$ can be mapped to separable states in $\tilde{\mathcal{S}}_{2}^{(N/2 +1)}$, as shown in Result~\ref{obs:sep}, not all separable states in $\mathcal{S}_{2}^{(N/2 +1)}$ lie in $\tilde{\mathcal{S}}_{2}^{(N/2 +1)}$.
    }
    \label{fig:mapping}
\end{figure}
%%%%%%%%%%%%%%%%%%%%%%%%%%%%%%%%%%%%%%%%%%%%%%%%%%%%%%%%%%%%%%%%
Furthermore, several theoretical techniques have been developed for symmetric states \cite{giraud2015tensor,liu2012local,bastin2009operational}, with a particular focus on the Majorana representation~\cite{aulbach2010maximally,martin2010multiqubit,markham2011entanglement,ribeiro2011entanglement}. Also, symmetric states have been characterized via their nonlocality~\cite{wang2012nonlocality} and spin non-classicality~\cite{ohst2025revealing}, as well as their single-qubit reductions~\cite{arnaud2013exploring,baguette2014multiqubit}. Yet, a comprehensive framework for symmetric entanglement remains elusive.

In this manuscript, we present a systematic approach to address $N$-qubit symmetric entanglement. Our idea is to map a multipartite state into a bipartite state of higher local dimensions and analyze the properties of the latter, as illustrated in Fig.~\ref{fig:mapping}. This mapping is based on an embedding between symmetric subspaces which was first pointed out in Ref.~\cite{toth2009entanglement}, and more recently considered in Refs.~\cite{marconi2023entanglement,romero2024multipartite}, in the context of the separability problem for multipartite symmetric states. Remarkably, our technique can be extended to high-dimensional mixed states (more details will appear elsewhere~\cite{inpreparation}).

We first recall the structure of the symmetric subspace, both in the multipartite and bipartite cases. Then, we present our mapping [Result~\ref{ob:mapping}] and prove that it preserves the separability of quantum states [Result~\ref{obs:sep}]. Also, we investigate how this mapping transforms the entanglement of a multipartite symmetric state in terms of the symmetric border rank [Result~\ref{res:border}], providing analytically computable lower bounds to the geometric measure [Result~\ref{ob:GM_LB}] and another entanglement measure [Results~\ref{ob:NewMeasure_LB}] based on the nuclear norm. Finally, we show that the complementary symmetric subspace to the mapped subspace is entangled [Result~\ref{ob:entangledsubspace}].

%%%%%%%%%%%%%%%%%%%%%%%%%%%%%%%%%%%%%%%%%%%%%%%%%%%%%%%%%%%%%%%%
\section{Symmetric subspace}
%\vspace{1em}
%{\it Symmetric subspace.---}
Let $\mathcal{H}_{N}^{(d)} = (\mathbb{C}^{d})^{\otimes N}$ be the $N$-qudit Hilbert space with dimension $d^N$, where each single-qudit Hilbert space $\mathbb{C}^{d}$ is spanned by the orthonormal bases $\{\ket{i}_{d}\}_{i=0}^{d-1}$. The symmetric subspace, denoted by $\mathcal{S}_{N}^{(d)}$, with dimension $\binom{N+d-1}{d-1}$, corresponds to the convex set spanned by the pure states that are invariant under any permutation of the parties.

For any $N \geq 2$ and $d=2$ (i.e., $N$ qubits), the subspace $\mathcal{S}_{N}^{(2)}$ has dimension $N+1$ and is spanned by the orthonormal Dicke states, i.e., $\{\ket{D^{k}_{N}}\}_{k=0}^{N}$, defined as
\begin{equation}
\label{dicke:qubits}
    \ket{D^{k}_{N}} = \binom{N}{k}^{-1/2} \sum_{\pi_{N}} \pi_{N} \left( \ket{0}^{\otimes(N-k)} \ket{1}^{\otimes k}\right)~,
\end{equation}
where $k$ denotes the number of excitations, and the sum runs over all possible permutations $\pi_{N}$ acting on $N$ qubits. Here and in the following, we omit the subscript for the qubit basis and set $\ket{0} \equiv \ket{0}_2$ and $\ket{1} \equiv \ket{1}_2$.  For any $d \geq 2$ and $N=2$ (i.e., two qudits), the subspace $\mathcal{S}_{2}^{(d)}$ has dimension $d(d+1)/2$ and is spanned by the orthonormal states $\{\ket{\psi^{(d)}_{ij}}\}_{i,j=0}^{d-1}$, given by
\begin{equation}
\label{dicke:twoqudits}
    \ket{\psi^{(d)}_{ii}} = \ket{ii}_{d}~,
    \quad
    \ket{\psi^{(d)}_{ij}} = \frac{\ket{ij}_{d} + \ket{ji}_{d}}{\sqrt{2}}~, \quad i\neq j,
\end{equation}
where $\ket{ij}_{d} = \ket{i}_{d} \otimes \ket{j}_{d}$. For example, the Greenberger-Horne-Zeilinger (GHZ) state in $N$ qubits, $\ket{{\rm GHZ}_N} \equiv (1/\sqrt{2}) (\ket{D^{0}_{N}} + \ket{D^{N}_{N}} )$, lives in the subspace $\mathcal{S}_{N}^{(2)}$, while the two-qudit maximally entangled state, $\ket{\psi_d^+} = (1/\sqrt{d}) \sum_{i=0}^{d-1} \ket{\psi^{(d)}_{ii}}$, lives in the subspace $\mathcal{S}_{2}^{(d)}$. 

%%%%%%%%%%%%%%%%%%%%%%%%%%%%%%%%%%%%%%%%%%%%%%%%%%%%%%%%%%%%%%%%
\section{Mapping}
%\vspace{1em}
%{\it Mapping.---}
Let us present a mapping that allows to cast an $N$-qubit symmetric state for any even $N$ as a high-dimensional bipartite symmetric state. 
\begin{result}\label{ob:mapping}
For any even $N$, there is a mapping $\mathcal{M}: \mathcal{S}_{N}^{(2)} \rightarrow \tilde{\mathcal{S}}_{2}^{(N/2 +1)}$ such that
\begin{equation}
\label{mapping}
    \mathcal{M}(\ket{D_{N}^{k}}) = \sum_{i \leq j=0}^{N/2}\delta_{k,i+j} \mu_{ij} \ket{\psi^{(N/2 +1)}_{ij}}~,
\end{equation}
where $\ket{D_{N}^{k}} \in \mathcal{S}_{N}^{(2)}$ is the Dicke state of Eq.~(\ref{dicke:qubits}), $\ket{\psi^{(N/2 +1)}_{ij}} \! \in \! \mathcal{S}_{2}^{(N/2 +1)}$ is the bipartite state of Eq.~(\ref{dicke:twoqudits}), and $\delta_{a,b}$ is the Kronecker delta. Here $\mu_{ij}$ is a factor given by
\begin{equation}
\label{mu}
\mu_{ii} = \frac{\binom{N/2}{i}}{\sqrt{\binom{N}{2i}}}~, \quad
\mu_{ij} = \sqrt{ 2\frac{\binom{N/2}{i} \binom{N/2}{j}}{\binom{N}{i+j}}}~,\quad i\neq j.
\end{equation}
The subspace $\tilde{\mathcal{S}}_{2}^{(N/2+1)}$ is spanned by the states $\{\mathcal{M}(\ket{D_{N}^{k}})\}_{k=0}^{N}$ with dimension $N+1$.
\end{result}
The proof is given in Appendix~\ref{ap:A_mapping}. We have three remarks on Result~\ref{ob:mapping}. First, this mapping does not correspond to an isomorphism between symmetric subspaces, but rather to an embedding, i.e.,
\begin{equation} \label{eq:inequality_tildeS}
    \tilde{\mathcal{S}}_{2}^{(N/2+1)} \subset \mathcal{S}_{2}^{(N/2+1)}~.
\end{equation}
This pictorial representation is illustrated in Fig.~\ref{fig:mapping}. Second, since any $\ket{\Psi} \in \mathcal{S}_{N}^{(2)}$ can be written as $\ket{\Psi} = \sum_{k} c_k \ket{D_N^k}$ for some coefficients $c_k$, one can always find the mapped state $\mathcal{M} (\ket{\Psi}) = \sum_{k} c_k \mathcal{M} (\ket{D_N^k}) \in \tilde{\mathcal{S}}_{2}^{(N/2 +1)}$. Finally, we prove in Appendix~\ref{ap:B_fidelity_preserve} that, for any $\ket{\Psi_1}, \ket{\Psi_2} \in \mathcal{S}_{N}^{(2)}$, the mapping $\mathcal{M}$ preserves the inner product, i.e.,
\begin{equation} 
\label{eq:fidelity_preserve}
    F(\ket{\Psi_1}, \ket{\Psi_2})
    = F[\mathcal{M}(\ket{\Psi_1}), \mathcal{M}(\ket{\Psi_2})]~,
\end{equation}
where $F(\ket{\Psi_1}, \ket{\Psi_2}) \equiv \lvert \braket{\Psi_1|\Psi_2} \rvert^2$ denotes the fidelity between two pure states. As a consequence, $\{\mathcal{M}(\ket{D^{k}_{N}})\}$ forms an orthonormal basis for the subspace $\tilde{\mathcal{S}}_{2}^{(N/2 +1)}$. 
    
%%%%%%%%%%%%%%%%%%%%%%%%%%%%%%%%%%%%%%%%%%%%%%%%%%%%%%%%%%%%%%%%
\section{Separability}
%\vspace{1em}
%{\it Separability.---}
Here we show that the mapping of Eq.~(\ref{mapping}) preserves the separability of quantum states, i.e., any separable state in $\mathcal{S}_{N}^{(2)}$ is mapped to a bipartite separable state in $\tilde{\mathcal{S}}_{2}^{(N/2+1)}$. Let us recall that a pure state $\ket{\Phi_{\rm sep}}\in \mathcal{H}_{N}^{(2)}$ is said (fully) separable if it can be cast as $\ket{\Phi_{\rm sep}} = \bigotimes_{i=1}^N \ket{\Phi_{i}}$ with $\ket{\Phi_{i}} \in \mathbb{C}^{2}$ for $i=1, \dots, N$. Otherwise, it is said entangled. In particular, a symmetric state is either genuinely multipartite entangled or fully separable~\cite{eckert2002quantum,ichikawa2008exchange,wei2010exchange}, i.e., $\ket{\Phi_{\rm sep}} = \ket{\Phi}^{\otimes N}$ with $\ket{\Phi}\in \mathbb{C}^{2}$. Hence, we find the following result:
\begin{result}
\label{obs:sep}
   Let $\ket{\Phi} \in \mathcal{S}_{N}^{(2)}$ be a multipartite state and $\mathcal{M}(\ket{\Phi}) \in \tilde{\mathcal{S}}_{2}^{(N/2+1)}$ be its mapped bipartite state. If $\ket{\Phi}$ is separable, then $\mathcal{M}(\ket{\Phi})$ is also separable. Conversely, if $\mathcal{M}(\ket{\Phi})$ is entangled, then $\ket{\Phi}$ is also entangled.
\end{result}
The proof is in Appendix~\ref{ap:C_proof_sep}. We stress that Result~\ref{obs:sep} does not imply that all separable states in $\mathcal{S}_{2}^{(N/2+1)}$ live in $\tilde{\mathcal{S}}_{2}^{(N/2+1)}$, as mentioned in Fig.~\ref{fig:mapping}. As an example, for $N=4$, the state $\ket{\psi^{(3)}_{11}}$ is separable, yet no symmetric separable state $\ket{\Phi_{\rm sep}} \in \mathcal{S}_{4}^{(2)}$ satisfies $\mathcal{M}(\ket{\Phi_{\rm sep}}) = \ket{\psi^{(3)}_{11}}$. Instead, $\ket{\psi^{(3)}_{11}} = \sqrt{{2}/{3}}\mathcal{M}(\ket{D_{4}^{2}}) + ({1}/{\sqrt{3}}) \ket{\hat{\psi}_{4}}$, where $\ket{\hat{\psi}_{4}} = ({1}/{\sqrt{3}}) \ket{\psi^{(3)}_{11}} -\sqrt{{2}/{3}}\ket{\psi^{(3)}_{02}}  \notin  \tilde{\mathcal{S}}^{(3)}_{2}$.

Remarkably, Result~\ref{obs:sep} can be generalized to mixed states. A mixed symmetric state $\vr \in \mathcal{S}_{N}^{(2)}$ can be expressed as $\vr = \sum_{i,j} \lambda_{ij} \ketbra{\Psi_i}{\Psi_j}$, with coefficients $\lambda_{ij}$ and $\ket{\Psi_i} \in \mathcal{S}_{N}^{(2)}$. The corresponding mixed-state mapping is defined as $\mathcal{M}(\vr) = \sum_{i,j} \lambda_{ij} \mathcal{M}(\ket{\Psi_i}) [\mathcal{M}(\ket{\Psi_i})]^\dagger$. Since any mixed separable state can be written as a convex combination of pure product states, we can conclude that for any $\vr_{\rm sep} \in \mathcal{S}_{N}^{(2)}$, the mapped state $\mathcal{M}(\vr_{\rm sep}) \in \tilde{\mathcal{S}}_{2}^{(N/2+1)}$ is also separable. Conversely, if $\mathcal{M}(\vr)$ is entangled, then $\vr$ is entangled. This means that the mapping allows for reducing the separability problem of multipartite symmetric states to the bipartite symmetric subspace. As an example, consider the mixed state $\vr_p = p \ketbra{{\rm W}_{N}}{{\rm W}_{N}} + (1-p) \Pi_{\mathcal{S}}/(N+1)$ with $\ket{{\rm W}_{N}} \equiv \ket{D_N^1}$ and $\Pi_{\mathcal{S}}$ being the projector onto $\mathcal{S}_{N}^{(2)}$ and set $N=6$. Applying the positive partial transpose (PPT) criterion~\cite{peres1996separability} to $\mathcal{M}({\vr_p})$, we find it is entangled for $p > 0.034$, which is smaller than the value $0.042$ presented in Ref.~\cite{devi2007characterizing}. Finally, we note that there exists a mapped bipartite state that is entangled but cannot be detected by the PPT criterion. Specifically, by applying our mapping to the four-qubit state in Eq.~(105) in \cite{tura2018separability} and to the four-qubit state $\rho_{BE4}$ in \cite{toth2009entanglement}, we obtain two-qutrit PPT-entangled states of the same form of the one reported in \cite{toth2009entanglement}.

%%%%%%%%%%%%%%%%%%%%%%%%%%%%%%%%%%%%%%%%%%%%%%%%%%%%%%%%%%%%%%%%
\section{Rank of a tensor}
%\vspace{1em}
%{\it Rank of a tensor.---}
Here we investigate the relation between the entanglement of a multipartite symmetric state and its mapped bipartite state. Let us recall that any pure bipartite state $\ket{\psi} \in  \mathcal{H}_{2}^{(d)}$ admits, up to local unitaries, a Schmidt decomposition of the form $\ket{\psi} = \sum_{i=1}^{r} \gamma_i \ket{a_i} \otimes \ket{b_i},$ where $\gamma_{i} \in \mathbb{R}^+$ and $\braket{a_i|a_j} = \braket{b_i|b_j} = \delta_{ij}$. The minimal number of terms, $r$, defines the \textit{Schmidt rank} of $\ket{\psi}$, denoted ${\rm R}(\ket{\psi})$~\cite{peres1997quantum,nielsen_chuang_2010}, which corresponds to the matrix rank of the reduced state $\vr_A = \tr_B(\ketbra{\psi}{\psi})$, and is thus analytically computable.

A generalization of the Schmidt rank for a multipartite state $\ket{\Psi} \in \mathcal{H}_{N}^{(d)}$, is given by the \textit{tensor rank}~\cite{dur2000three,eisert2001schmidt,chitambar2008tripartite}, denoted ${\rm T}(\ket{\Psi})$, which corresponds to the minimal $r$ such that $\ket{\Psi} = \sum_{i=1}^{r} t_i \ket{\Psi_i^{(1)}} \otimes \cdots \otimes \ket{\Psi_i^{(N)}}$. Here, $t_i \in \mathbb{C}$ and the states $\ket{\Psi_i^{(k)}}$ are not necessarily orthogonal. Examples for three qubits are $T(\ket{{\rm GHZ}_3}) = 2$ and $T(\ket{{\rm W}_3}) = 3$. Unlike the Schmidt rank, computing the tensor rank is, in general, an NP-hard problem~\cite{haastad1990tensor}. Another related concept is the \textit{border rank}~\cite{bini1980approximate,landsberg2011tensors,allman2013tensor,vrana2015asymptotic,zuiddam2017note}, denoted ${\rm B}(\ket{\Psi})$, which corresponds to the minimum $r$ such that $\ket{\Psi}$ can be approximated arbitrarily closely by a sequence $\{\ket{\Psi_{\epsilon}}\}$ with ${\rm T}(\ket{\Psi_{\epsilon}})= r$ for all $\epsilon \neq 0$. Clearly, ${\rm B}(\ket{\Psi}) \leq {\rm T}(\ket{\Psi})$. A notable example is given by
\begin{equation}
    \ket{{\rm W}_3} = \lim_{\epsilon \rightarrow 0}
    \frac{1}{\epsilon}\left[(\ket{0} + \epsilon \ket{1})^{\otimes 3} -  \ket{0}^{\otimes 3}\right],
\end{equation}
which shows that ${\rm B}(\ket{{\rm W}_3}) = 2$ (see Review~\cite{bruzda2024rank}).

For a symmetric state $\ket{\Psi}\in \mathcal{S}_{N}^{(d)}$, one may look for symmetric decompositions and introduce alternative notions of ranks. The \textit{symmetric tensor rank}~\cite{comon2008symmetric,landsberg2010ranks}, denoted ${\rm ST}(\ket{\Psi})$, corresponds to the minimal $r$ such that $\ket{\Psi} = \sum_{i=1}^{r} s_i \ket{\Psi_i}^{\otimes N}$ for $s_i \in \mathbb{C}$. Similarly, the \textit{symmetric border rank}, denoted ${\rm SB}(\ket{\Psi})$, is defined through approximated symmetric decompositions, satisfying ${\rm SB}(\ket{\Psi}) \leq {\rm ST}(\ket{\Psi})$. By construction, one has ${\rm T}(\ket{\Psi}) \leq {\rm ST}(\ket{\Psi})$ and ${\rm B}(\ket{\Psi}) \leq {\rm SB}(\ket{\Psi})$.

It has been conjectured that ${\rm T}(\ket{\Psi}) = {\rm ST}(\ket{\Psi})$ holds for any symmetric state, widely known as Comon's conjecture~\cite{comon2008symmetric}. An analogous conjecture exists for the border rank, asserting ${\rm B}(\ket{\Psi}) = {\rm SB}(\ket{\Psi})$~\cite{buczynski2013determinantal,friedland2016remarks}. Despite several efforts in this area (see \cite{zhang2016comon} and references therein), a conclusive result remains elusive. Here we establish the following result:
\begin{result}
\label{res:border}
    Let $\ket{\Psi} \in \mathcal{S}_{N}^{(2)}$ be a multipartite state. Then,
    \begin{equation}
         {\rm R}[\mathcal{M}(\ket{\Psi})] ={\rm SB}(\ket{\Psi}).
    \end{equation}
\end{result}
The proof is given in Appendix~\ref{ap:D_border}, where we used tools from algebraic geometry, exploiting the correspondence between symmetric states and homogeneous polynomials. Our result provides not only an analytical calculation of the symmetric border rank, but it also establishes a connection to the Schmidt rank of the mapped state. Explicit computations for several states are given in Appendix~\ref{ap:E_examples_tensor_rank}.
 
%%%%%%%%%%%%%%%%%%%%%%%%%%%%%%%%%%%%%%%%%%%%%%%%%%%%%%%%%%%%%%%%
\section{Geometric measure}
%\vspace{1em}
%{\it Geometric measure.---}
Here we show that the amount of multipartite symmetric entanglement can be analytically estimated via the mapping. Consider the geometric measure of entanglement for a state $\ket{\Psi} \in \mathcal{H}_{N}^{(d)}$~\cite{shimony1995degree,wei2003geometric,weinbrenner2025quantifying}:
\begin{equation} \label{eq:def_gm}
    E(\ket{\Psi}) =
    1 - \Lambda^2(\ket{\Psi}),
\end{equation}
where $\Lambda(\ket{\Psi}) \equiv \max_{\ket{\Phi_{\rm sep}} \in \mathcal{H}_{N}^{(d)}} \sqrt{F(\ket{\Phi_{\rm sep}}, \ket{\Psi})}$ is called the spectral norm (see Review~\cite{bruzda2024rank}), and the maximization is taken over all product states. For any bipartite state $\ket{\psi} \in \mathcal{H}_{2}^{(d)}$, it holds that $\Lambda^2(\ket{\psi}) = \max_i \gamma_i^2 \equiv \gamma_{\max}^2$, where $\gamma_i$ is the Schmidt coefficient of $\ket{\psi}$~\cite{bourennane2004experimental}. For a symmetric state $\ket{\Psi} \in \mathcal{S}_{N}^{(d)}$, the geometric measure can be efficiently computed by restricting to symmetric product states~\cite{hubener2009geometric}, i.e.,
\begin{equation}\label{eq:lemma_fidelity}
    \Lambda(\ket{\Psi})
    = \Lambda_{\rm sym}(\ket{\Psi}),
\end{equation}
where $\Lambda_{\rm sym}(\ket{\Psi})$ is defined by replacing $\ket{\Phi_{\rm sep}} \in \mathcal{H}_{N}^{(d)}$ in $\Lambda(\ket{\Psi})$ with $\ket{\Phi_{\rm sep}} \in \mathcal{S}_{N}^{(d)}$.

Now we can present the following result:
\begin{result}\label{ob:GM_LB}
    Let $\ket{\Psi} \in \mathcal{S}_{N}^{(2)}$ be a multipartite state. Then,
    \begin{equation} \label{eq:ob_GM_inequality}
         1- \gamma_{\max}^2 \leq E(\ket{\Psi}),
    \end{equation}
    where $\gamma_{\max}$ is the maximal Schmidt coefficient of the mapped state $\mathcal{M}(\ket{\Psi})$.
\end{result}

\begin{proof}
   We note that $E[\mathcal{M}(\ket{\Psi})] \leq E(\ket{\Psi})$ is equivalent to $\Lambda_{\rm sym}^2(\ket{\Psi}) \leq \Lambda_{\rm sym}^2[\mathcal{M}(\ket{\Psi})]$, according to Eqs.~(\ref{eq:def_gm}, \ref{eq:lemma_fidelity}). This can be shown as
    \begin{subequations}
        \begin{align}
        \label{eq:derivation1_GM_LB}
        \Lambda_{\rm sym}^2(\ket{\Psi})
        \!&=\! \! \!
        \max_{\mathcal{M}(\ket{\Phi_{\rm sep}}) \in \tilde{\mathcal{S}}_{2}^{(N/2 + 1)}} 
        \! \! \! \! \! \! \! \! \!
        F \left[\mathcal{M}(\ket{\Phi_{\rm sep}}), \mathcal{M}(\ket{\Psi}) \right]\\
        \label{eq:derivation2_GM_LB}
        &\leq \max_{\ket{\phi_{\rm sep}} \in \mathcal{S}_{2}^{(N/2 + 1)}} 
        F \left[ \ket{\phi_{\rm sep}}, \mathcal{M}(\ket{\Psi}) \right]\\
        \label{eq:derivation3_GM_LB}
        &=\Lambda_{\rm sym}^2[\mathcal{M}(\ket{\Psi})],
    \end{align}
    \end{subequations}
    where in Eq.~(\ref{eq:derivation1_GM_LB}) we used Eq.~(\ref{eq:fidelity_preserve}), and in Eq.~(\ref{eq:derivation2_GM_LB}), we employed Eq.~(\ref{eq:inequality_tildeS}). Thus, the proof is completed.
\end{proof}

We have several remarks. First, Result~\ref{ob:GM_LB} provides an analytical estimate of the geometric measure for multipartite symmetric states. In particular, it can be computed exactly when the inequality (\ref{eq:ob_GM_inequality}) is saturated, i.e., when the closest product state lies in $\tilde{\mathcal{S}}_{2}^{(N/2 + 1)}$, such as $\ket{0}^{\otimes N}$. In fact, the GHZ state achieves this saturation.

Next, we collect some cases where, for a symmetric entangled state $\ket{\Psi}$ with large $E(\ket{\Psi})$, its mapped state $\mathcal{M}(\ket{\Psi})$ also retains a large value $E[\mathcal{M}(\ket{\Psi})]$. In particular, for the maximally entangled symmetric (MES) states for $N=4,6$~\cite{martin2010multiqubit,aulbach2010maximally} (also see \cite{aulbach2012classification}), given by 
\begin{subequations}
    \begin{align} \label{eq:MES_4}
        \ket{\Psi_4^{\rm MES}} &= \sqrt{\tfrac{1}{3}} \ket{D_4^0} + \sqrt{\tfrac{2}{3}} \ket{D_4^3},
        \\
        \ket{\Psi_6^{\rm MES}} &= \sqrt{\tfrac{1}{2}} \ket{D_6^1} + \sqrt{\tfrac{1}{2}} \ket{D_6^5},
        \label{eq:MES_6}
    \end{align}
\end{subequations}
the mapped states $\mathcal{M}(\ket{\Psi_4^{\rm MES}})$ and $\mathcal{M}(\ket{\Psi_6^{\rm MES}})$ remain maximally entangled. On the other hand, for $N\!=\!8$, the candidate MES state, $\ket{\Psi_8^{\rm MES}} \!=\! \alpha \ket{D_8^1} \!+\! \beta \ket{D_8^6}$, with $\alpha \! \approx \! 0.672$ and $\beta \! \approx \! 0.741$, as numerically found in Ref.~\cite{aulbach2010maximally}, does not become maximally entangled after the mapping, despite achieving a high value $E[\mathcal{M}(\ket{\Psi_8^{\rm MES}})] \!=\! 0.816$. 

Finally, we note that the mapping does not preserve the ordering of entangled states in general:
\begin{equation}
    E(\ket{\Psi_1}) \geq E(\ket{\Psi_2})
    \! \nRightarrow \!
    E[\mathcal{M}(\ket{\Psi_1})] \geq E[\mathcal{M}(\ket{\Psi_2})],
\end{equation}
for symmetric states $\ket{\Psi_1}, \ket{\Psi_2}$. Examples are $\ket{D_N^2}$ and $\ket{{\rm W}_N}$ for $N \geq 4$, satisfying $E(\ket{D_N^2}) = 1-2 (N-1) (N-2)^{N-2}/N^{N-1} > E(\ket{{\rm W}_N}) = 1- [(N-1)/N]^{N-1}$~\cite{wei2003geometric}, but $E[\mathcal{M}(\ket{D_N^2})] = 1-N/[2 (N-1)] < E[\mathcal{M}(\ket{{\rm W}_N})] = 1/2$. Beyond the geometric measure, finding multipartite entanglement measures that preserve the ordering for all two symmetric entangled states under the mapping is left for future work.

%%%%%%%%%%%%%%%%%%%%%%%%%%%%%%%%%%%%%%%%%%%%%%%%%%%%%%%%%%%%%%%%
\section{Entanglement from nuclear norm}
%\vspace{1em}
%{\it Entanglement from nuclear norm.---}
Let us consider another entanglement measure for a state $\ket{\Psi} \in \mathcal{H}_{N}^{(d)}$~\cite{derksen2017theoretical}:
\begin{equation}
    H(\ket{\Psi}) = 1 - \frac{1}{\Xi^2(\ket{\Psi})},
\end{equation}
where $\Xi(\ket{\Psi})$ is the nuclear norm~\cite{friedland2018nuclear}, defined as
\begin{equation} \label{eq:defofnuclearnorm}
    \Xi(\ket{\Psi}) \!=\! 
    \min \bigg\{ \!
    \sum_{i=1}^r |\xi_i|,
    \, \mbox{s.t.} \,
    \ket{\Psi} \!=\! \! \sum_{i=1}^r \xi_i
    \bigotimes_{k=1}^N \ket{\psi_{i}^{(k)}} 
    \! \bigg\},
\end{equation}
where $\xi_i \in \mathbb{C}$ and $\ket{\psi_{i}^{(k)}}$ are not necessarily orthogonal.

The nuclear norm is the dual norm of the spectral norm~\cite{lim2013blind}, i.e., $\Xi(\ket{\Psi}) = \max \{ \lvert \braket{\Psi|X} \rvert, \, {\rm s.t.} \, \Lambda(\ket{X}) \leq 1 \}$, which leads to the inequality $1 \leq \Xi(\ket{\Psi}) \Lambda(\ket{\Psi})$. As a consequence, we have $E(\ket{\Psi}) \leq H(\ket{\Psi})$. For any bipartite state $\ket{\psi} \in \mathcal{H}_{2}^{(d)}$, it holds that $\Xi(\ket{\psi}) = \sum_{i=1}^r \gamma_i$ for its Schmidt coefficient $\gamma_i$, which is connected to the negativity of $\ket{\psi}$ given by $\mathcal{N}(\ket{\psi}) = (1/2) [(\sum_{i=1}^r \gamma_i)^2 -1 ]$~\cite{vidal2002computable}. Using our mapping, we prove the analogue of Result~\ref{ob:GM_LB} for the entanglement measure $H(\ket{\Psi})$ of a symmetric state:
\begin{result}\label{ob:NewMeasure_LB}
    Let $\ket{\Psi} \in \mathcal{S}_{N}^{(2)}$ be a multipartite state. Then,
    \begin{equation} \label{eq:ob_NewMeasure_inequality}
         1- \frac{1}{2 \mathcal{N} + 1} \leq H(\ket{\Psi}),
    \end{equation}
    where $\mathcal{N}$ is the negativity of the mapped state $\mathcal{M}(\ket{\Psi})$.
\end{result}
\begin{proof}
    We note that for a symmetric state $\ket{\Psi} \in \mathcal{S}_{N}^{(d)}$, the following relation holds~\cite{friedland2018nuclear}:
    \begin{equation}\label{eq:lemma_nuclear_norm}
    \Xi(\ket{\Psi})
    = \Xi_{\rm sym}(\ket{\Psi}),
    \end{equation}
    where $\Xi_{\rm sym}(\ket{\Psi})$ is defined by replacing $\bigotimes_{k=1}^N \ket{\psi_{i}^{(k)}} $ in Eq.~(\ref{eq:defofnuclearnorm}) with $\ket{\psi_i}^{\otimes N}$. Since this is the analogue of Eq.~(\ref{eq:lemma_fidelity}) and the nuclear norm is the dual norm of the spectral norm, one can show that $\Xi_{\rm sym}^2(\ket{\Psi}) \leq \Xi_{\rm sym}^2[\mathcal{M}(\ket{\Psi})]$, similarly to the proof of Result~\ref{ob:GM_LB}. Thus, the proof is completed.
\end{proof}

Based on this, we can derive a necessary condition for a state to be a MES with respect to the geometric measure. For an MES state $\ket{\Psi_N^{\rm MES}}$, it holds that $H [\mathcal{M}(\ket{\Psi_N^{\rm MES}})] \leq E (\ket{\Psi_N^{\rm MES}})$. Otherwise, the state is not MES. This can be shown by using the fact that $E(\ket{\Psi_N^{\rm MES}}) = H(\ket{\Psi_N^{\rm MES}})$ holds for MES states $\ket{\Psi_N^{\rm MES}}$~\cite{derksen2017theoretical}. In fact, we can check that the candidate MES state $\ket{\Psi_8^{\rm MES}}$ has $H[\mathcal{M}(\ket{\Psi_8^{\rm MES}})] = 0.793$ and $E (\ket{\Psi_8^{\rm MES}}) = 0.816$.

%%%%%%%%%%%%%%%%%%%%%%%%%%%%%%%%%%%%%%%%%%%%%%%%%%%%%%%%%%%%%%%%
\section{Entangled symmetric subspace}
%\vspace{1em}
%{\it Entangled symmetric subspace.---}
Finally, using our approach, we reveal the existence of an entangled subspace in the bipartite symmetric subspace. Let us recall that a subspace is called entangled if all pure states in it are entangled~\cite{parthasarathy2004maximal,bhat2006completely}. A famous example is the bipartite antisymmetric subspace, denoted as $\mathcal{A}_{2}^{(d)}$, with dimension $d(d-1)/2$~\cite{parthasarathy2004maximal}, where $\mathcal{H}_2^{(d)} \!=\! \mathcal{S}_{2}^{(d)} \oplus \mathcal{A}_{2}^{(d)}$ and $\mathcal{A}_{2}^{(d)}$ is spanned by the orthonormal basis $\ket{\chi^{(d)}_{ij}} \!=\! (1/\sqrt{2})(\ket{ij}_{d} - \ket{ji}_{d})$. It is interesting to ask whether there is a nontrivial entangled subspace within $\mathcal{S}_{2}^{(d)}$.

We answer this question affirmatively as follows:
\begin{result} \label{ob:entangledsubspace}
\label{obs:subspace}
   Let $\mathcal{S}_{2}^{(d)}$ be the two-qudit symmetric subspace and $\tilde{\mathcal{S}}_{2}^{(d)}$ be its mapped subspace, where $d = N/2 +1$ for any even $N$. Then, the complementary subspace $\hat{\mathcal{S}}_{2}^{(d)}$, such that $\mathcal{S}_{2}^{(d)} =  \tilde{\mathcal{S}}_{2}^{(d)} \oplus \hat{\mathcal{S}}_{2}^{(d)}$, is entangled with dimension $(d-1)(d-2)/2$ (or, equivalently, $N(N-2)/8$).
\end{result}
The proof is given in Appendix~\ref{ap:F_entangledsubspace}, and the subspace $\hat{\mathcal{S}}_{2}^{(d)}$ corresponds to the darkest orange-colored region in Fig.~\ref{fig:mapping}. There we will show that for any $\ket{\psi_{\hat{\mathcal{S}}}} \in \hat{\mathcal{S}}_{2}^{(d)}$, the geometric measure is strictly greater than zero. Result~\ref{ob:entangledsubspace} provides a fundamental consequence: in the two-qudit Hilbert space, $\mathcal{H}_2^{(d)} = \tilde{\mathcal{S}}_{2}^{(d)} \oplus \hat{\mathcal{S}}_{2}^{(d)} \oplus \mathcal{A}_{2}^{(d)}$, all states in $\hat{\mathcal{S}}_{2}^{(d)}$ and $\mathcal{A}_{2}^{(d)}$ are entangled for any $d$. The simplest case is $d=3$ (i.e., $N = 4$), when $\hat{\mathcal{S}}_{2}^{(3)}$ is a one-dimensional subspace spanned by $\ket{\hat{\psi}_{4}} = ({1}/{\sqrt{3}}) \ket{\psi^{(3)}_{11}} -\sqrt{{2}/{3}}\ket{\psi^{(3)}_{02}}$. 

%%%%%%%%%%%%%%%%%%%%%%%%%%%%%%%%%%%%%%%%%%%%%%%%%%%%%%%%%%%%%%%%
\section{Conclusion}
%\vspace{1em}
%{\it Conclusion.---}
We presented a systematic approach to map multipartite symmetric states onto bipartite symmetric states of higher local dimension. This mapping allows for the analysis of the original multipartite entangled state via its mapped counterpart. Finally, we uncovered the existence of nontrivial entangled symmetric subspaces, whose states are not reached by the mapping. 

Our findings open several avenues for further research. 
First, it would be worthwhile to relate the entanglement criterion discussed after Result~\ref{obs:sep} to the PPT criterion~\cite{peres1996separability}. In particular, it would be interesting to investigate whether the mapping preserves the PPT condition. Second, connecting our results to the Majorana representation~\cite{aulbach2010maximally,martin2010multiqubit} may provide further insights into the structure of MES states. Also, it would be interesting to use the Schmidt rank to explore the entanglement dimensionality in the complementary subspace $\hat{\mathcal{S}}_{2}^{(d)}$~\cite{cubitt2008dimension}. More broadly, our results may encourage advances in Bell nonlocality, alternative entanglement measures, and might be extended beyond the symmetric subspace. Finally, an experimental implementation of the proposed mapping would be of great interest. To this end, platforms based on atomic ensembles---such as those exploited in \cite{kunkel2018spatially,fadel2018spatial,lange2018entanglement}---appear particularly promising for future experiments.

%%%%%%%%%%%%%%%%%%%%%%%%%%%%%%%%%%%%%%%%%%%%%%%%%%%%%%%%%%%%%%%
\section*{Acknowledgments}
%\vspace{1em}
%{\it Acknowledgments.---}
We would like to thank Otfried G\"uhne and Anna Sanpera for discussions. In particular, we are very grateful to John Martin for pointing out an error in Version 1 of this manuscript. We also thank Shmuel Friedland for informing us on Eq.~(\ref{eq:lemma_nuclear_norm}) and Giorgio Ottaviani for clarifying several mathematical details in Appendix~\ref{ap:D_border}.
CM acknowledges support from the European Union - NextGeneration EU, "Integrated infrastructure initiative in Photonic and Quantum Sciences" - I-PHOQS [IR0000016, ID D2B8D520, CUP B53C22001750006]. 
SI acknowledges support from Horizon Europe programme HORIZON-CL4-2022-QUANTUM-02-SGA via the project 101113690 (PASQuanS2.1) and JST ASPIRE (JPMJAP2339).

%%%%%%%%%%%%%%%%%%%%%%%%%%%%%%%%%%%%%%%%%%%%%%%%%%%%
\onecolumngrid
\appendix

%%%%%%%%%%%%%%%%%%%%%%%%%%%%%%%%%%%%%%%%%%%%%%%%%%%%%%%%%%%%%%%
\section{Proof of Result~\ref{ob:mapping}} \label{ap:A_mapping}
Here we give the proof of Result~\ref{ob:mapping} presented in the main text by showing that
\begin{equation} \label{eq:ap:map1}
    \mathcal{M}(\ket{D_{N}^{k}}) = \sum_{i \leq j=0}^{N/2}\delta_{k,i+j} \mu_{ij} \ket{\psi^{(N/2 +1)}_{ij}},
    \quad
    \mu_{ii} = \frac{\binom{N/2}{i}}{\sqrt{\binom{N}{2i}}},
    \quad
    \mu_{ij} = \sqrt{ 2\frac{\binom{N/2}{i} \binom{N/2}{j}}{\binom{N}{i+j}}},
\end{equation}
where $\delta_{a,b}$ is the Kronecker-delta symbol and $\ket{\psi^{(N/2 +1)}_{ij}} \in \mathcal{S}_{2}^{(N/2 +1)}$.

\begin{proof}
We start by recalling that any $N$-qubit Dicke state can be decomposed as \cite{stockton2003characterizing,toth2007detection,raveh2024dicke}
\begin{equation}
\label{eq:dickedecomposition}
    \ket{D^{k}_{N}}
    = \sum_{i=\max\{0,k-N/2\}}^{\min\{k,N/2\}}
    \sqrt{\frac{\binom{N/2}{i} \binom{N/2}{k-i}}{\binom{N}{k}}}
    \ket{D_{N/2}^{i}} \! \ket{D_{N/2}^{k-i}}~.
\end{equation}
Assuming $k\geq N/2$ and using the fact that $\binom{N/2}{k-i}=0$ for $k-i> N/2$, Eq.~(\ref{eq:dickedecomposition}) takes the form
\begin{equation}
     \label{eq:dickedecomposition1b}
     \ket{D_{N}^{k}}=\sum_{i=0}^{N/2} \sqrt{\frac{\binom{N/2}{i} \binom{N/2}{k-i}}{\binom{N}{k}}} \ket{D_{N/2}^{i}} \ket{D_{N/2}^{k-i}}.
\end{equation}
Recalling that $\mathcal{S}_{N}^{(2)} \cong \mathbb{C}^{N+1}$, we can set $\ket{D_{N/2}^{i}} \equiv \ket{i}_{N/2+1}$. Hence, Eq.~(\ref{eq:dickedecomposition1b}) can be rewritten as
\begin{equation}
\label{eq:convenient}
     \mathcal{M}(\ket{D^{k}_{N}}) =
     \sum_{i,j=0}^{N/2}  \delta_{k,i+j} \sqrt{\frac{\binom{N/2}{i} \binom{N/2}{j}}{\binom{N}{k}}} \ket{i}_{N/2+1} \ket{j}_{N/2+1}~.
\end{equation}
Splitting the sum into $i = j$ and $i \neq j$ allows to identify $\mu_{ij}$ as in Eq.~(\ref{eq:ap:map1}), thus leading to the conclusion. An analogous reasoning yields the same result when~$k < N/2$.
\end{proof}

For the sake of clarity, we here provide an explicit example of the mapping by considering the four-qubit Dicke state $\ket{D^{1}_{4}} = (1/2)(\ket{1000} +\ket{0100} +\ket{0010} +\ket{0001})$. The first step is to split it into two parts:
\begin{equation}
    \ket{D^{1}_{4}}
    \label{dicke:4qubits:split}
    = \frac{1}{\sqrt{2}}\left[\left(\frac{\ket{10}
    + \ket{01}}{\sqrt{2}}\right) \! \ket{00}
    + \ket{00} \! \left(\frac{\ket{10}
    + \ket{01}}{\sqrt{2}}\right)\right].
\end{equation}
Next, Eq.~(\ref{dicke:4qubits:split}) can be rewritten in terms of the two-qubit states $\ket{D^{k}_{2}}$, i.e., 
\begin{equation}
\label{dicke:4qubits:map}
    \ket{D^{1}_{4}} = \frac{1}{\sqrt{2}}\left(\ket{D^{1}_{2}} \ket{D^{0}_{2}} + \ket{D^{0}_{2}} \ket{D^{1}_{2}}\right)~,
\end{equation}
where 
$\ket{D^{0}_{2}} = \ket{00}$,
$\ket{D^{1}_{2}} = (1/\sqrt{2})(\ket{10}+\ket{01})$, and
$\ket{D^{2}_{2}} = \ket{11}$.
Finally, notice that Eq.~(\ref{dicke:4qubits:map}) can be cast as
\begin{equation}
    \ket{D^{1}_{4}} = \frac{1}{\sqrt{2}}\left(\ket{1}_{3}\ket{0}_{3} + \ket{0}_{3}\ket{1}_{3}\right) \equiv \ket{\psi^{(3)}_{01}}~,
\end{equation}
where we have set $\ket{D^{k}_{2}} = \ket{k}_{3}$ for $k=0,1,2$, using the fact that $\mathcal{S}^{(2)}_{2}$ has dimension $3$, i.e., $\mathcal{S}^{(2)}_{2} \cong \mathbb{C}^{3}$. Consequently, any Dicke state $\ket{D^{k}_{4}}$ can be expressed as a linear combination of the two-qutrit states $\ket{\psi^{(3)}_{ij}}$ as follows:
\begin{equation}
    \ket{D^{0}_{4}} = \ket{\psi^{(3)}_{00}},
    \quad
    \ket{D^{1}_{4}} = \ket{\psi^{(3)}_{01}},
    \quad
    \ket{D^{2}_{4}} = \frac{1}{\sqrt{3}}(\ket{\psi^{(3)}_{02}}+\sqrt{2}\ket{\psi^{(3)}_{11}} ),
    \quad
    \ket{D^{3}_{4}} = \ket{\psi^{(3)}_{12}},
    \quad
    \ket{D^{4}_{4}} = \ket{\psi^{(3)}_{22}}.
\end{equation}

%%%%%%%%%%%%%%%%%%%%%%%%%%%%%%%%%%%%%%%%%%%%%%%%%%%%%%%%%%%%%%%
\section{Proof of Eq.~(\ref{eq:fidelity_preserve})} \label{ap:B_fidelity_preserve}
Here we give the proof of Eq.~(\ref{eq:fidelity_preserve}) presented in the main text by showing that for any $\ket{\psi_1}, \ket{\psi_2} \in \mathcal{S}_{N}^{(2)}$, it holds that
\begin{equation} 
    F(\ket{\psi_1}, \ket{\psi_2})
    = F[\mathcal{M}(\ket{\psi_1}), \mathcal{M}(\ket{\psi_2})],
\end{equation}
where $F(\ket{\psi_1}, \ket{\psi_2}) \equiv \lvert \braket{\psi_1|\psi_2} \rvert^2$.

\begin{proof}
    We begin by writing $\ket{\psi_1} = \sum_{k} a_k \ket{D_N^k}$ and $\ket{\psi_2} = \sum_{k} b_k \ket{D_N^k}$ with coefficients $a_k, b_k \in \mathbb{C}$. Using Result~\ref{ob:mapping} in the main text, we have that
    \begin{equation}
        \mathcal{M} (\ket{\psi_1})
        = \sum_{k} \sum_{i \leq j=0}^{N/2} a_k \delta_{k,i+j} \mu_{ij} \ket{\psi^{(N/2 +1)}_{ij}},
        \quad
        \mathcal{M} (\ket{\psi_2})
        = \sum_{k} \sum_{i \leq j=0}^{N/2} b_k \delta_{k,i+j} \mu_{ij} \ket{\psi^{(N/2 +1)}_{ij}}.
    \end{equation}
    Therefore, $F[\mathcal{M}(\ket{\psi_1}), \mathcal{M}(\ket{\psi_2})]$ can be given by
    \begin{subequations}
    \begin{align}
    F[\mathcal{M}(\ket{\psi_1}), \mathcal{M}(\ket{\psi_2})]
    &= \left|
    \sum_{k,k^\prime} \sum_{i \leq j=0}^{N/2} \sum_{i^\prime \leq j^\prime =0}^{N/2}
    a_k^* b_{k^\prime}
    \delta_{k,i+j} \delta_{k^\prime,i^\prime + j^\prime}
    \mu_{ij} \mu_{i^\prime j^\prime}
    \braket{\psi^{(N/2 +1)}_{ij}|\psi^{(N/2 +1)}_{i^\prime j^\prime}}
    \right|^2
    \\
    &= \left|
    \sum_{k,k^\prime} \sum_{i \leq j=0}^{N/2} \sum_{i^\prime \leq j^\prime =0}^{N/2}
    a_k^* b_{k^\prime}
    \delta_{k,i+j} \delta_{k^\prime,i^\prime + j^\prime}
    \mu_{ij} \mu_{i^\prime j^\prime}
    \delta_{i i^\prime} \delta_{j j^\prime}
    \right|^2
    \\
    &= \left|
    \sum_{k} \sum_{i \leq j=0}^{N/2}
    a_k^* b_{k}
    \delta_{k,i+j}
    \mu_{ij}^2
    \right|^2
    \\
    &= \left|
    \sum_{k} a_k^* b_{k} f(k,N)
    \right|^2,
    \quad
    f(k,N) \equiv \sum_{i \leq j=0}^{N/2} \delta_{k,i+j} \mu_{ij}^2.
    \end{align}
    \end{subequations}
    Then, it is sufficient to show $f(k,N) = 1$ for any $0 \leq k \leq N$. Splitting the sum in $f(k,N)$ into $i = j$ and $i \neq j$ yields
    \begin{subequations}
    \begin{align}
        f(k,N) &= 
        \sum_{i=0}^{N/2} \delta_{k,2i} \mu_{ii}^2
        + \frac{1}{2} \sum_{i,j=0, i \neq j}^{N/2} \delta_{k,i+j} \mu_{ij}^2
        \\
        &=  \sum_{i=0}^{N/2} \delta_{k,2i} \left(\frac{\binom{N/2}{i}}{\sqrt{\binom{N}{2i}}}\right)^2
        + \sum_{i,j=0, i \neq j}^{N/2} \delta_{k,i+j} \frac{\binom{N/2}{i} \binom{N/2}{j}}{\binom{N}{i+j}}
        \\
        &=\sum_{i,j=0}^{N/2} \delta_{k,i+j}  \frac{\binom{N/2}{i} \binom{N/2}{j}}{\binom{N}{i+j}}
        \\
         &=\frac{1}{\binom{N}{k}} \sum_{i=0}^{N/2}  \binom{N/2}{i} \binom{N/2}{k-i}
         =\frac{1}{\binom{N}{k}} \sum_{i=0}^{k}  \binom{N/2}{i} \binom{N/2}{k-i}
         = 1~,
    \end{align}
    \end{subequations}
    where in the last line we have used the fact that $\binom{N/2}{k-i}=0$ for any $i > k$, and the Chu-Vandermonde identity, i.e.,
    \begin{equation}
    \sum_{i=0}^{k}  \binom{N/2}{i} \binom{N/2}{k-i} = \binom{N}{k}~.
    \end{equation}
    Hence, we complete the proof.
\end{proof}

%%%%%%%%%%%%%%%%%%%%%%%%%%%%%%%%%%%%%%%%%%%%%%%%%%%%%%%%%%%%%%%
\section{Proof of Result~\ref{obs:sep}} \label{ap:C_proof_sep}
Here we give the proof of Result~\ref{obs:sep} presented in the main text by showing that for an $N$-qubit separable symmetric state $\ket{\Phi_{\rm sep}} \in \mathcal{S}_{N}^{(2)}$, the mapped bipartite state $\mathcal{M}(\ket{\Phi_{\rm sep}}) \in \tilde{\mathcal{S}}_{2}^{(N/2+1)}$ is separable.

\begin{proof}
    Let us begin by noticing that any $N$-qubit symmetric separable state can be cast as
    \begin{equation}
     \ket{\Phi_{\rm sep}}
     = \ket{\Phi}^{\otimes N}
     = \sum_{k=0}^{N} \sqrt{\binom{N}{k}}\Phi_{0}^{N-k} \Phi_{1}^{k} \ket{D_{N}^{k}}~,
     \end{equation}
     where $\ket{\Phi}=\sum_{i=0}^{1}\Phi_{i}\ket{i} \in \mathbb{C}^{2}$. Applying the mapping of Eq.~(\ref{mapping}) in the main text yields
     \begin{equation}
     \label{eq:mapped_sep}
     \mathcal{M}(\ket{\Phi_{\rm sep}})
     = \sum_{i \leq j=0}^{N/2} \sqrt{\binom{N}{i+j}} \Phi_{0}^{N-i-j} \Phi_{1}^{i+j} \mu_{ij} \ket{\psi^{(N/2+1)}_{ij}}~.
    \end{equation}
    Recall that a two-qudit symmetric separable state can be written as
    \begin{equation}
    \ket{\phi_{\rm sep}}
    = \ket{\phi}^{\otimes 2}
    = \sum_{i=0}^{d-1} \phi_{i}^{2} \ket{\psi^{(d)}_{ii}}
    + \sum_{i<j=0}^{d-1} \sqrt{2} \phi_{i}\phi_{j} \ket{\psi^{(d)}_{ij}}~,
    \end{equation}
    with $\ket{\phi}=\sum_{i=0}^{d-1}\phi_{i}\ket{i} \in \mathbb{C}^{d}$. Setting $\phi_{i} = \sqrt{\binom{N/2}{i}}\Phi_{0}^{N/2-i} \Phi_{1}^{i}$, an explicit calculation yields $\mathcal{M}(\ket{\Phi_{\rm sep}}) = \ket{\phi_{\rm sep}}$, thus concluding the proof.
\end{proof}

%%%%%%%%%%%%%%%%%%%%%%%%%%%%%%%%%%%%%%%%%%%%%%%%%%%%%%%%%%%%%%%%%%
\section{Proof of Result~\ref{res:border}} \label{ap:D_border}
\noindent Here we prove Result~\ref{res:border} in the main text by showing that $ {\rm R}[\mathcal{M}(\ket{\Psi})] = {\rm SB}(\ket{\Psi})$.
\begin{proof}
In what follows, we will make use of some tools from algebraic geometry, where a key role is played by multivariate polynomials. Similarly to the case of quantum states, also for polynomials several ranks can be introduced. For instance, the \textit{polynomial rank} of $p(x_{0},x_{1}, \dots, x_{d-1})$, denoted ${\rm PR}(p)$, is defined as the minimum number $r$ such that there exists a decomposition of the form $p(x_{0},x_{1}, \dots, x_{d-1}) = \sum_{i=1}^{r} (x_{0} \beta_{0,i} + \dots x_{0} \beta_{d-1,i})^{N}$ \cite{chen2010tensor}. Similarly, the symmetric border rank ${\rm SB} (p)$ of a polynomial is defined as the minimum integer $r$ such that there exists a sequence of homogeneous polynomials $\{p_{\epsilon}\}$ with polynomial rank ${\rm PR} (p) = r$ and such that $\lim_{\epsilon \rightarrow \infty} p_{\epsilon} = p$.
Remarkably, there exists an isomorphism between a state $\ket{\Psi} \in \mathcal{S}_{N}^{(d)}$ and a homogeneous polynomial $p(x_{0},x_{1}, \dots, x_{d-1}) $ of degree $N$ in $d$ variables (see, e.g., \cite{chen2010tensor,bernardi2012algebraic}). 
In particular, a monomial $x_{0}^{k_{0}} x_{1}^{k_{1}} \dots x_{d-1}^{k_{d-1}}$ is associated to a quantum state with $k_{0}$ systems in $\ket{0}$, $k_{1}$ systems in $\ket{1}$ and so on. As an example, focusing on the qubit case (i.e., $d=2$), the 4-qubit state $\ket{0101}$ corresponds to $x_{0}^{2}x_{1}^2$.
As a consequence, a state $\ket{\Psi} = \sum_{k=0}^{N}\alpha_{k} \ket{D_{N}^{k}}$ can be associated to a polynomial $p_{\Psi}$ of the form 
\begin{equation}
\label{dicke_poly}
  p_{\Psi} = \sum_{k=0}^{N} \binom{N}{k} \frac{\alpha_{k}}{\sqrt{\binom{N}{k}}}x_{0}^{N-k} x_{1}^{k}~.
\end{equation}
Leveraging on this correspondence, the symmetric tensor rank of a quantum state can be identified with the polynomial rank of its associated polynomial, i.e., ${\rm ST}(\ket{\Psi}) \equiv {\rm PR}(p_{\Psi})$ \cite{chen2010tensor}. Analogously, we have ${\rm SB}(\ket{
\Psi}) \equiv {\rm SB}(p_{\Psi})$ \cite{bernardi2012algebraic}. 

A celebrated result in algebraic geometry is the Sylvester's algorithm \cite{sylvester1886extension} (see also \cite{brachat2010symmetric, comas2011rank} for some modern versions), which allows to compute the polynomial rank of a bivariate polynomial $p(x_{0},x_1)$. In combination with a theorem by Comas and Seiguer \cite{comas2011rank}, a simplified algorithm can be obtained which provides also its symmetric border rank (see Algorithm 2.62 in \cite{bernardi2018hitchhiker}). 
In order to illustrate this result, let us introduce the $n$-th \textit{catalecticant matrix} $C_{N-n,n}(p)$ \cite{kanev1999chordal} of a polynomial $p(x_{0},x_1) = \sum_{k=0}^{N}\binom{N}{k}a_{k} x_{0}^{N-k}x_{1}^{k}$. This is defined as the $(N-n+1)\times (n+1)$ matrix with elements $[C_{N-n,n}(p)]_{ij} = a_{i+j}$, with $i=0, \dots, N-n$ and $j=0, \dots, n$. 
Hence, we have the following \cite{bernardi2018hitchhiker}:
\begin{equation}
\label{eq:sylvester}
     rank(C_{N/2,N/2}(p)) = {\rm SB (p)}~.
\end{equation}

We want to prove that
\begin{equation}
\label{eq:rank_cat}
rank(C_{\Psi}) = {\rm R [\mathcal{M}(\ket{\Psi})]}~,
\end{equation}
where we have denoted $C_{\Psi}\equiv C_{N/2,N/2}(p_{\Psi})$ the $N/2$-catalecticant matrix associated to a state $\ket{\Psi} \in \mathcal{S}_{N}^{(2)}$.

Let us now consider a state $\ket{\Psi} = \sum_{k=0}^{N} \alpha_{k} \ket{D_{N}^{k}}$. As a consequence of Eq.~(\ref{dicke_poly}), the catalecticant matrix $C_{\Psi}$ has elements $[C_{\Psi}]_{ij} = \alpha_{i+j}/\sqrt{\binom{N}{i+j}}$.

Turning to the mapped state $\mathcal{M}(\ket{\Psi})$, its Schmidt rank corresponds to the matrix rank of the reduced density matrix $\varrho_{\mathcal{M}(\ket{\Psi)}} = \tr_{B}[\mathcal{M}(\ket{\Psi}) \mathcal{M}(\ket{\Psi})^{\dagger}]$. First, note that we can write
\begin{align}
\varrho_{\mathcal{M}(\ket{\Psi)}} &= \sum_{i,j,k=0}^{N/2} \alpha_{i+j} \alpha_{k+j} \lambda_{ij} \lambda_{jk} \ketbra{i}{k} \\
&= \sum_{i,k=0}^{N/2} \sum_{j=0}^{N/2}A_{ij} A_{jk} \ketbra{i}{k} = A A^{T}~,
\end{align}
where we have introduced a matrix $A$ such that
\begin{equation}
    \quad A_{ij} = \alpha_{i+j} \lambda_{ij}~, \quad \lambda_{ij} = \sqrt{\frac{\binom{N/2}{i}\binom{N/2}{j}}{\binom{N}{i+j}}}~.
\end{equation}
As a consequence,
\begin{equation}
    {\rm R [\mathcal{M}(\ket{\Psi})]} = rank(\varrho_{\mathcal{M}(\ket{\Psi)}}) = rank (A A^{T}) = rank(A)~.
\end{equation}
A direct calculation shows that we can write
\begin{equation}
    A = D C_{\Psi}D^{T}~,
\end{equation}
where $D$ is a diagonal matrix with elements $D_{ii} = \sqrt{\binom{N/2}{i}}$. Since $D$ is a full-rank matrix, it follows that
\begin{equation}
    rank(A) = rank(D C_{\Psi}D^{T}) = rank(C_{\Psi})~,
\end{equation}
from which we find Eq.~(\ref{eq:rank_cat}). Then, using Eq.~(\ref{eq:sylvester}), the proof is completed.
\end{proof}

%%%%%%%%%%%%%%%%%%%%%%%%%%%%%%%%%%%%%%%%%%%%%%%%%%%%%%%%%%%%%%%%%%
\section{Computation of the symmetric border rank of some multipartite symmetric states}\label{ap:E_examples_tensor_rank}
Here we compute explicitly the border rank of some symmetric states of $N$ qubits. We start by showing that ${\rm SB}(\ket{D_{N}^{k}})$ is given by $k+1$ for $k\leq N/2$ and $N-k+1$ for $k> N/2$. 
\begin{proof}
Using Result~\ref{res:border}, to compute the border rank of $\ket{D_{N}^{k}}$ is enough to calculate the Schmidt rank of the state $\mathcal{M}(\ket{D_{N}^{k}})$. This is equivalent to compute the matrix rank of the reduced state $\vr_{\ket{D_{N}^{k}}}= \tr_{B}[\mathcal{M}(\ket{D_{N}^{k}})\mathcal{M}(\ket{D_{N}^{k}})^{\dagger}]$, where $\tr_{B}$ denotes the partial trace over the second party. Note that, due to the permutational invariance, the expression of $\vr_{\ket{D_{N}^{k}}}$ does not depend on the choice of the party. Using Eq.~(\ref{eq:convenient}) in Appendix~\ref{ap:A_mapping}, we can write
\begin{align}
    \vr_{\ket{D_{N}^{k}}} &= \sum_{i, j=0}^{N/2} \sum_{m, n=0}^{N/2} \delta_{i+j,k} \delta_{m+n,k} \sqrt{\frac{\binom{N/2}{i}\binom{N/2}{j}}{\binom{N}{k}}} \sqrt{\frac{\binom{N/2}{m}\binom{N/2}{n}}{\binom{N}{k}}} \tr_{B}[\ketbra{ij}{mn}]~,\\
    &= \sum_{i,j,n=0}^{N/2} \delta_{i+j,k} \delta_{m+j,k} \sqrt{\binom{N/2}{i}\binom{N/2}{m}} \frac{\binom{N/2}{j}}{\binom{N}{k}} \ketbra{i}{m}~,\\
    &= \sum_{i,j=0}^{N/2} \delta_{i+j,k} \frac{\binom{N/2}{i} \binom{N/2}{j}}{\binom{N}{k}} \ketbra{i}{i}~,\\
    \label{eq:dicke_red_state}
    &= \sum_{i=0}^{N/2} \frac{\binom{N/2}{i} \binom{N/2}{k-i}}{\binom{N}{k}} \ketbra{i}{i}~.
\end{align}

\noindent Since $\vr_{k}$ is diagonal in the computational basis, its matrix rank corresponds to the number of non-zero elements. This can be easily computed observing that the last term in Eq.~(\ref{eq:dicke_red_state}) can be cast as (see also Eq.~(\ref{eq:dickedecomposition}))
\begin{equation}
   \vr_{\ket{D_{N}^{k}}} = \sum_{i=\max\{0,k-N/2\}}^{\min\{k,N/2\}} \frac{\binom{N/2}{i}\binom{N/2}{k-i}}{\binom{N}{k}}\ketbra{i}{i}~.
\end{equation}
Hence, when $k\leq N/2$, the matrix rank of $\vr_{\ket{D_{N}^{k}}}$ is found to be $k+1$. Analogously, when $k > N/2$, the matrix rank is $ N-k+1$. Hence, the proof is complete.
\end{proof}

As a consequence of this result, since the $N$-qubit W and GHZ states can be expressed as linear combinations of Dicke states, it is immediate to compute their border rank for any even $N$. Indeed, we have $\ket{{\rm W}_{N}} \equiv \ket{D_{N}^{1}}$, from which it follows ${\rm SB}(\ket{{\rm W}_{N}})= 2$. 

Similarly, considering the GHZ state, we can set $\ket{{\rm GHZ}_{N}} \equiv (\ket{D^{0}_{N}}+ \ket{D^{N}_{N}})/\sqrt{2}$, from which we have ${\rm SB}(\ket{{\rm GHZ}_{N}}) = 2$. The above results can be summarized in the following table:
\begin{table}[h] 
    \centering
\[
\begin{array}{|c|c|c|c|}
\hline
\ket{\Psi} & \ket{{\rm W}_{N}} & \ket{{\rm GHZ}_{N}}  & \ket{D_{N}^{k}} \\
\hline
{\rm SB}(\ket{\Psi}) & 2 & 2 &  \begin{array}{c}
k+1, ~k\leq N/2  \\ 
N-k+1, ~k > N/2 \\
\end{array}\\
\hline
\end{array}
\]
\caption{Symmetric border rank of some $N$-qubit symmetric states.}
    \label{tab:mappedranks}
\end{table}

%%%%%%%%%%%%%%%%%%%%%%%%%%%%%%%%%%%%%%%%%%%%%%%%%%%%%%%
\section{Proof of Result~\ref{ob:entangledsubspace}} \label{ap:F_entangledsubspace}
Here we give the proof of Result~\ref{ob:entangledsubspace} presented in the main text by showing that for any state $\ket{\psi_{\hat{\mathcal{S}}}} \in \hat{\mathcal{S}}_{2}^{(d)}$, the geometric measure is nonzero, i.e., $E(\ket{\psi_{\hat{\mathcal{S}}}}) > 0$.

\begin{proof}
    We begin by denoting $\Pi_{\mathcal{S}}$, $\Pi_{\tilde{\mathcal{S}}}$, and $\Pi_{\hat{\mathcal{S}}}$ as the projectors onto the subspaces ${\mathcal{S}}_{2}^{(d)}$, $\tilde{\mathcal{S}}_{2}^{(d)}$, and $\hat{\mathcal{S}}_{2}^{(d)}$, respectively. Note that $\Pi_{\mathcal{S}} = \Pi_{\tilde{\mathcal{S}}} + \Pi_{\hat{\mathcal{S}}} = (1/2) (\eins + \swap)$, where $\swap$ denotes the SWAP operator such that $\swap \ket{ab} = \ket{ba}$. Recalling that $E(\ket{\psi}) = 1- \max_{\ket{ab}} \, \lvert \braket{ab|\psi} \rvert^2$ and $\hat{\mathcal{S}}_{2}^{(d)}$ is the symmetric subspace, the relation in Eq.~(\ref{eq:lemma_fidelity}) in the main text reduces our task to show that $\max_{\ket{a}} \, \lvert \braket{aa|\psi_{\hat{\mathcal{S}}}} \rvert^2 < 1$ for $\ket{\psi_{\hat{\mathcal{S}}}} \in \hat{\mathcal{S}}_{2}^{(d)}$.
    
    Using Eq.~(40) in Ref.~\cite{pankowski2010few}, we have
    \begin{equation}
        \max_{\ket{a}} \, \lvert \braket{aa|\psi_{\hat{\mathcal{S}}}} \rvert^2
        \leq \max_{\ket{a}} \, \braket{aa|\Pi_{\hat{\mathcal{S}}}|aa}
        = 1 - \min_{\ket{a}} \, \braket{aa|\Pi_{\tilde{\mathcal{S}}}|aa},
    \end{equation}
    where in the second equality, we used the relation that $\Pi_{\hat{\mathcal{S}}} = \Pi_{\mathcal{S}} - \Pi_{\tilde{\mathcal{S}}}$. This yields
    \begin{equation}
        E(\ket{\psi_{\hat{\mathcal{S}}}})
        \geq
        \min_{\ket{a}} \, \braket{aa|\Pi_{\tilde{\mathcal{S}}}|aa}.
    \end{equation}    
    Then our task is further reduced to show that $\min_{\ket{a}} \, \braket{aa|\Pi_{\tilde{\mathcal{S}}}|aa} > 0$. Letting $\ket{a} = \sum_{i=0}^{d-1} a_i \ket{i}$, we write
    \begin{equation}
        \ket{a}^{\otimes 2}
        = \left(\sum_{i=0}^{d-1} a_i \ket{i} \right)^{\otimes 2}
        = \sum_{i} a_i^2 \ket{\psi^{(d)}_{ii}}
        + \frac{1}{2} \sum_{i \neq j} \sqrt{2} a_i a_j \ket{\psi^{(d)}_{ij}}.
    \end{equation}
    Inserting this form into $\braket{aa|\Pi_{\tilde{\mathcal{S}}}|aa}$ and using $\Pi_{\tilde{\mathcal{S}}} = \sum_{k=0}^N \ket{\phi_k}\! \bra{\phi_k}$ with $\ket{\phi_k} = \sum_{i \leq j=0}^{d-1}\delta_{k,i+j} \mu_{ij} \ket{\psi^{(d)}_{ij}}$, we obtain
    \begin{align}
        \braket{aa|\Pi_{\tilde{\mathcal{S}}}|aa}
        &= \sum_{k=0}^N \lvert \braket{aa|\phi_k} \rvert^2
        \\
        &= \sum_{k=0}^N 
        \left| \sum_{i=0}^{d-1} \delta_{k,2i} \mu_{ii} a_i^2
        + \frac{\sqrt{2}}{4} \sum_{i \neq j} \sum_{l \neq m} \delta_{k,l+m} \mu_{lm}
        a_i a_j (\delta_{il} \delta_{jm} + \delta_{im} \delta_{jl})        
        \right|^2
        \\
        &= \sum_{k=0}^N 
        \left| \sum_{i=0}^{d-1} \delta_{k,2i} \mu_{ii} a_i^2
        + \frac{\sqrt{2}}{2} \sum_{i \neq j} \delta_{k,i+j} \mu_{ij} a_i a_j        
        \right|^2
        \\
        &\equiv \sum_{k=0}^N 
        \left| \sum_{i,j=0}^{d-1} a_i M_{ij}^{(k)} a_j
        \right|^2,
    \end{align}
    where we denoted the matrix $M^{(k)} = (M_{ij}^{(k)})$ with elements
    \begin{equation} \label{eq:def_mat_M}
        M_{ii}^{(k)} = \delta_{k,2i} \mu_{ii},
        \quad
        M_{ij}^{(k)} = \frac{\sqrt{2}}{2} \delta_{k,i+j} \mu_{ij}.
    \end{equation}
    Furthermore, letting $\vec{a} = (a_0, \ldots, a_{d-1})^\top \in \mathbb{C}^{d}$, we have
    \begin{equation}
        \sum_{k=0}^N 
        \left| \sum_{i,j=0}^{d-1} a_i M_{ij}^{(k)} a_j
        \right|^2
        = \sum_{k=0}^N 
        \left| \vec{a}^\top M^{(k)} \vec{a} \right|^2.
    \end{equation}

    Thus, our task is reduced as follows:
    \begin{equation}
        E(\ket{\psi_{\hat{\mathcal{S}}}})
        \geq 
        \min_{\ket{a}} \, \braket{aa|\Pi_{\tilde{\mathcal{S}}}|aa}
        = 
        \min_{\vec{a}} \, \left\{ \sum_{k=0}^N \left| \vec{a}^\top M^{(k)} \vec{a} \right|^2 \right\}
        \geq
        \sum_{k=0}^N \left\{ \min_{\vec{a}} \, \left| \vec{a}^\top M^{(k)} \vec{a} \right|^2 \right\}
        = 
        \sum_{k=0}^N \{\sigma_{\rm min} [M^{(k)} ]\}^2,
    \end{equation}
    where $\sigma_{\rm min}[M^{(k)}]$ denotes the minimal absolute eigenvalue of $M^{(k)}$. From the definition of the matrix $M^{(k)}$ in Eq.~(\ref{eq:def_mat_M}), we notice that for $k<N/2$ or $k>N/2$, it can be always written as the following block form
    \begin{equation}
        M^{(k<N/2)} =
        \begin{bmatrix}
        A_k & 0 \\
        0 & 0
        \end{bmatrix},
        \quad
        M^{(k>N/2)} =
        \begin{bmatrix}
        0 & 0 \\
        0 & A_k
        \end{bmatrix},
    \end{equation}
    for some matrices $A_k$. In this case, the matrix $M^{(k)}$ is not full-rank, i.e., $\sigma_{\rm min}[M^{(k)}] = 0$. On the other hand, for $k=N/2$, the matrix $M^{(k=N/2)}$ can be written as the anti-diagonal form, where $M_{ij}^{(k=N/2)} \neq 0$ for $i+j = N/2$, but $M_{ij}^{(k=N/2)} = 0$ for all the other $i,j$. Then the matrix $M^{(k=N/2)}$ is full-rank, i.e., $\sigma_{\rm min}[M^{(k=N/2)}] > 0$. Hence we can complete the proof.
\end{proof}

%%%%%%%%%%%%%%%%%%%%%%%%%%%%%%%%%%%%%%%%%%%%%%%%%%%%%%%%%%%%%%%

%\bibliographystyle{apsrev4-1}
%\bibliography{ref.bib}

\begin{thebibliography}{79}%
\makeatletter
\providecommand \@ifxundefined [1]{%
 \@ifx{#1\undefined}
}%
\providecommand \@ifnum [1]{%
 \ifnum #1\expandafter \@firstoftwo
 \else \expandafter \@secondoftwo
 \fi
}%
\providecommand \@ifx [1]{%
 \ifx #1\expandafter \@firstoftwo
 \else \expandafter \@secondoftwo
 \fi
}%
\providecommand \natexlab [1]{#1}%
\providecommand \enquote  [1]{``#1''}%
\providecommand \bibnamefont  [1]{#1}%
\providecommand \bibfnamefont [1]{#1}%
\providecommand \citenamefont [1]{#1}%
\providecommand \href@noop [0]{\@secondoftwo}%
\providecommand \href [0]{\begingroup \@sanitize@url \@href}%
\providecommand \@href[1]{\@@startlink{#1}\@@href}%
\providecommand \@@href[1]{\endgroup#1\@@endlink}%
\providecommand \@sanitize@url [0]{\catcode `\\12\catcode `\$12\catcode `\&12\catcode `\#12\catcode `\^12\catcode `\_12\catcode `\%12\relax}%
\providecommand \@@startlink[1]{}%
\providecommand \@@endlink[0]{}%
\providecommand \url  [0]{\begingroup\@sanitize@url \@url }%
\providecommand \@url [1]{\endgroup\@href {#1}{\urlprefix }}%
\providecommand \urlprefix  [0]{URL }%
\providecommand \Eprint [0]{\href }%
\providecommand \doibase [0]{http://dx.doi.org/}%
\providecommand \selectlanguage [0]{\@gobble}%
\providecommand \bibinfo  [0]{\@secondoftwo}%
\providecommand \bibfield  [0]{\@secondoftwo}%
\providecommand \translation [1]{[#1]}%
\providecommand \BibitemOpen [0]{}%
\providecommand \bibitemStop [0]{}%
\providecommand \bibitemNoStop [0]{.\EOS\space}%
\providecommand \EOS [0]{\spacefactor3000\relax}%
\providecommand \BibitemShut  [1]{\csname bibitem#1\endcsname}%
\let\auto@bib@innerbib\@empty
%</preamble>
\bibitem [{\citenamefont {Ekert}(1991)}]{ekert1991quantum}%
  \BibitemOpen
  \bibfield  {author} {\bibinfo {author} {\bibfnamefont {A.~K.}\ \bibnamefont {Ekert}},\ }\href {\doibase 10.1103/PhysRevLett.67.661} {\bibfield  {journal} {\bibinfo  {journal} {Physical Review Letters}\ }\textbf {\bibinfo {volume} {67}},\ \bibinfo {pages} {661} (\bibinfo {year} {1991})}\BibitemShut {NoStop}%
\bibitem [{\citenamefont {Curty}\ \emph {et~al.}(2004)\citenamefont {Curty}, \citenamefont {Lewenstein},\ and\ \citenamefont {L{\"u}tkenhaus}}]{curty2004entanglement}%
  \BibitemOpen
  \bibfield  {author} {\bibinfo {author} {\bibfnamefont {M.}~\bibnamefont {Curty}}, \bibinfo {author} {\bibfnamefont {M.}~\bibnamefont {Lewenstein}}, \ and\ \bibinfo {author} {\bibfnamefont {N.}~\bibnamefont {L{\"u}tkenhaus}},\ }\href {\doibase 10.1103/PhysRevLett.92.217903} {\bibfield  {journal} {\bibinfo  {journal} {Physical Review Letters}\ }\textbf {\bibinfo {volume} {92}},\ \bibinfo {pages} {217903} (\bibinfo {year} {2004})}\BibitemShut {NoStop}%
\bibitem [{\citenamefont {Pezz{\'e}}\ and\ \citenamefont {Smerzi}(2009)}]{pezze2009entanglement}%
  \BibitemOpen
  \bibfield  {author} {\bibinfo {author} {\bibfnamefont {L.}~\bibnamefont {Pezz{\'e}}}\ and\ \bibinfo {author} {\bibfnamefont {A.}~\bibnamefont {Smerzi}},\ }\href {\doibase 10.1103/PhysRevLett.102.100401} {\bibfield  {journal} {\bibinfo  {journal} {Physical Review Letters}\ }\textbf {\bibinfo {volume} {102}},\ \bibinfo {pages} {100401} (\bibinfo {year} {2009})}\BibitemShut {NoStop}%
\bibitem [{\citenamefont {Raussendorf}\ and\ \citenamefont {Briegel}(2001)}]{raussendorf2001one}%
  \BibitemOpen
  \bibfield  {author} {\bibinfo {author} {\bibfnamefont {R.}~\bibnamefont {Raussendorf}}\ and\ \bibinfo {author} {\bibfnamefont {H.~J.}\ \bibnamefont {Briegel}},\ }\href {\doibase 10.1103/PhysRevLett.86.5188} {\bibfield  {journal} {\bibinfo  {journal} {Physical Review Letters}\ }\textbf {\bibinfo {volume} {86}},\ \bibinfo {pages} {5188} (\bibinfo {year} {2001})}\BibitemShut {NoStop}%
\bibitem [{\citenamefont {Horodecki}\ \emph {et~al.}(2009)\citenamefont {Horodecki}, \citenamefont {Horodecki}, \citenamefont {Horodecki},\ and\ \citenamefont {Horodecki}}]{horodecki2009quantum}%
  \BibitemOpen
  \bibfield  {author} {\bibinfo {author} {\bibfnamefont {R.}~\bibnamefont {Horodecki}}, \bibinfo {author} {\bibfnamefont {P.}~\bibnamefont {Horodecki}}, \bibinfo {author} {\bibfnamefont {M.}~\bibnamefont {Horodecki}}, \ and\ \bibinfo {author} {\bibfnamefont {K.}~\bibnamefont {Horodecki}},\ }\href {\doibase 10.1103/RevModPhys.81.865} {\bibfield  {journal} {\bibinfo  {journal} {Reviews of Modern Physics}\ }\textbf {\bibinfo {volume} {81}},\ \bibinfo {pages} {865} (\bibinfo {year} {2009})}\BibitemShut {NoStop}%
\bibitem [{\citenamefont {G{\"u}hne}\ and\ \citenamefont {T{\'o}th}(2009)}]{guhne2009entanglement}%
  \BibitemOpen
  \bibfield  {author} {\bibinfo {author} {\bibfnamefont {O.}~\bibnamefont {G{\"u}hne}}\ and\ \bibinfo {author} {\bibfnamefont {G.}~\bibnamefont {T{\'o}th}},\ }\href {\doibase 10.1016/j.physrep.2009.02.004} {\bibfield  {journal} {\bibinfo  {journal} {Physics Reports}\ }\textbf {\bibinfo {volume} {474}},\ \bibinfo {pages} {1} (\bibinfo {year} {2009})}\BibitemShut {NoStop}%
\bibitem [{\citenamefont {Friis}\ \emph {et~al.}(2019)\citenamefont {Friis}, \citenamefont {Vitagliano}, \citenamefont {Malik},\ and\ \citenamefont {Huber}}]{friis2019entanglement}%
  \BibitemOpen
  \bibfield  {author} {\bibinfo {author} {\bibfnamefont {N.}~\bibnamefont {Friis}}, \bibinfo {author} {\bibfnamefont {G.}~\bibnamefont {Vitagliano}}, \bibinfo {author} {\bibfnamefont {M.}~\bibnamefont {Malik}}, \ and\ \bibinfo {author} {\bibfnamefont {M.}~\bibnamefont {Huber}},\ }\href {\doibase https://doi.org/10.1038/s42254-018-0003-5} {\bibfield  {journal} {\bibinfo  {journal} {Nature Reviews Physics}\ }\textbf {\bibinfo {volume} {1}},\ \bibinfo {pages} {72} (\bibinfo {year} {2019})}\BibitemShut {NoStop}%
\bibitem [{\citenamefont {Erhard}\ \emph {et~al.}(2020)\citenamefont {Erhard}, \citenamefont {Krenn},\ and\ \citenamefont {Zeilinger}}]{erhard2020advances}%
  \BibitemOpen
  \bibfield  {author} {\bibinfo {author} {\bibfnamefont {M.}~\bibnamefont {Erhard}}, \bibinfo {author} {\bibfnamefont {M.}~\bibnamefont {Krenn}}, \ and\ \bibinfo {author} {\bibfnamefont {A.}~\bibnamefont {Zeilinger}},\ }\href {\doibase https://doi.org/10.1038/s42254-020-0193-5} {\bibfield  {journal} {\bibinfo  {journal} {Nature Reviews Physics}\ }\textbf {\bibinfo {volume} {2}},\ \bibinfo {pages} {365} (\bibinfo {year} {2020})}\BibitemShut {NoStop}%
\bibitem [{\citenamefont {Peres}(1995)}]{peres1995higher}%
  \BibitemOpen
  \bibfield  {author} {\bibinfo {author} {\bibfnamefont {A.}~\bibnamefont {Peres}},\ }\href {\doibase https://doi.org/10.1016/0375-9601(95)00315-T} {\bibfield  {journal} {\bibinfo  {journal} {Physics Letters A}\ }\textbf {\bibinfo {volume} {202}},\ \bibinfo {pages} {16} (\bibinfo {year} {1995})}\BibitemShut {NoStop}%
\bibitem [{\citenamefont {Plenio}\ and\ \citenamefont {Virmani}(2014)}]{plenio2005introduction}%
  \BibitemOpen
  \bibfield  {author} {\bibinfo {author} {\bibfnamefont {M.~B.}\ \bibnamefont {Plenio}}\ and\ \bibinfo {author} {\bibfnamefont {S.~S.}\ \bibnamefont {Virmani}},\ }\href {\doibase 10.1007/978-3-319-04063-9_8} {\bibfield  {journal} {\bibinfo  {journal} {Quantum Information and Coherence}\ ,\ \bibinfo {pages} {173}} (\bibinfo {year} {2014})}\BibitemShut {NoStop}%
\bibitem [{\citenamefont {Harrow}(2013)}]{harrow2013church}%
  \BibitemOpen
  \bibfield  {author} {\bibinfo {author} {\bibfnamefont {A.~W.}\ \bibnamefont {Harrow}},\ }\href {\doibase 10.48550/arXiv.1308.6595} {\bibfield  {journal} {\bibinfo  {journal} {arXiv preprint}\ } (\bibinfo {year} {2013}),\ 10.48550/arXiv.1308.6595}\BibitemShut {NoStop}%
\bibitem [{\citenamefont {T{\'o}th}\ and\ \citenamefont {G{\"u}hne}(2009)}]{toth2009entanglement}%
  \BibitemOpen
  \bibfield  {author} {\bibinfo {author} {\bibfnamefont {G.}~\bibnamefont {T{\'o}th}}\ and\ \bibinfo {author} {\bibfnamefont {O.}~\bibnamefont {G{\"u}hne}},\ }\href {\doibase 10.1103/PhysRevLett.102.170503} {\bibfield  {journal} {\bibinfo  {journal} {Physical Review Letters}\ }\textbf {\bibinfo {volume} {102}},\ \bibinfo {pages} {170503} (\bibinfo {year} {2009})}\BibitemShut {NoStop}%
\bibitem [{\citenamefont {Eisert}\ and\ \citenamefont {Briegel}(2001)}]{eisert2001schmidt}%
  \BibitemOpen
  \bibfield  {author} {\bibinfo {author} {\bibfnamefont {J.}~\bibnamefont {Eisert}}\ and\ \bibinfo {author} {\bibfnamefont {H.~J.}\ \bibnamefont {Briegel}},\ }\href {\doibase 10.1103/PhysRevA.64.022306} {\bibfield  {journal} {\bibinfo  {journal} {Physical Review A}\ }\textbf {\bibinfo {volume} {64}},\ \bibinfo {pages} {022306} (\bibinfo {year} {2001})}\BibitemShut {NoStop}%
\bibitem [{\citenamefont {Chitambar}\ \emph {et~al.}(2008)\citenamefont {Chitambar}, \citenamefont {Duan},\ and\ \citenamefont {Shi}}]{chitambar2008tripartite}%
  \BibitemOpen
  \bibfield  {author} {\bibinfo {author} {\bibfnamefont {E.}~\bibnamefont {Chitambar}}, \bibinfo {author} {\bibfnamefont {R.}~\bibnamefont {Duan}}, \ and\ \bibinfo {author} {\bibfnamefont {Y.}~\bibnamefont {Shi}},\ }\href {\doibase 10.1103/PhysRevLett.101.140502} {\bibfield  {journal} {\bibinfo  {journal} {Physical Review Letters}\ }\textbf {\bibinfo {volume} {101}},\ \bibinfo {pages} {140502} (\bibinfo {year} {2008})}\BibitemShut {NoStop}%
\bibitem [{\citenamefont {D{\"u}r}\ \emph {et~al.}(2000)\citenamefont {D{\"u}r}, \citenamefont {Vidal},\ and\ \citenamefont {Cirac}}]{dur2000three}%
  \BibitemOpen
  \bibfield  {author} {\bibinfo {author} {\bibfnamefont {W.}~\bibnamefont {D{\"u}r}}, \bibinfo {author} {\bibfnamefont {G.}~\bibnamefont {Vidal}}, \ and\ \bibinfo {author} {\bibfnamefont {J.~I.}\ \bibnamefont {Cirac}},\ }\href {\doibase 10.1103/PhysRevA.62.062314} {\bibfield  {journal} {\bibinfo  {journal} {Physical Review A}\ }\textbf {\bibinfo {volume} {62}},\ \bibinfo {pages} {062314} (\bibinfo {year} {2000})}\BibitemShut {NoStop}%
\bibitem [{\citenamefont {Lo}\ and\ \citenamefont {Popescu}(2001)}]{lo2001concentrating}%
  \BibitemOpen
  \bibfield  {author} {\bibinfo {author} {\bibfnamefont {H.-K.}\ \bibnamefont {Lo}}\ and\ \bibinfo {author} {\bibfnamefont {S.}~\bibnamefont {Popescu}},\ }\href {\doibase 10.1103/PhysRevA.63.022301} {\bibfield  {journal} {\bibinfo  {journal} {Physical Review A}\ }\textbf {\bibinfo {volume} {63}},\ \bibinfo {pages} {022301} (\bibinfo {year} {2001})}\BibitemShut {NoStop}%
\bibitem [{\citenamefont {H{\aa}stad}(1990)}]{haastad1990tensor}%
  \BibitemOpen
  \bibfield  {author} {\bibinfo {author} {\bibfnamefont {J.}~\bibnamefont {H{\aa}stad}},\ }\href {\doibase 10.1016/0196-6774(90)90014-} {\bibfield  {journal} {\bibinfo  {journal} {Journal of algorithms}\ }\textbf {\bibinfo {volume} {11}},\ \bibinfo {pages} {644} (\bibinfo {year} {1990})}\BibitemShut {NoStop}%
\bibitem [{\citenamefont {Chen}\ \emph {et~al.}(2010)\citenamefont {Chen}, \citenamefont {Chitambar}, \citenamefont {Duan}, \citenamefont {Ji},\ and\ \citenamefont {Winter}}]{chen2010tensor}%
  \BibitemOpen
  \bibfield  {author} {\bibinfo {author} {\bibfnamefont {L.}~\bibnamefont {Chen}}, \bibinfo {author} {\bibfnamefont {E.}~\bibnamefont {Chitambar}}, \bibinfo {author} {\bibfnamefont {R.}~\bibnamefont {Duan}}, \bibinfo {author} {\bibfnamefont {Z.}~\bibnamefont {Ji}}, \ and\ \bibinfo {author} {\bibfnamefont {A.}~\bibnamefont {Winter}},\ }\href {\doibase 10.1103/PhysRevLett.105.200501} {\bibfield  {journal} {\bibinfo  {journal} {Physical Review Letters}\ }\textbf {\bibinfo {volume} {105}},\ \bibinfo {pages} {200501} (\bibinfo {year} {2010})}\BibitemShut {NoStop}%
\bibitem [{\citenamefont {Comon}\ \emph {et~al.}(2008)\citenamefont {Comon}, \citenamefont {Golub}, \citenamefont {Lim},\ and\ \citenamefont {Mourrain}}]{comon2008symmetric}%
  \BibitemOpen
  \bibfield  {author} {\bibinfo {author} {\bibfnamefont {P.}~\bibnamefont {Comon}}, \bibinfo {author} {\bibfnamefont {G.}~\bibnamefont {Golub}}, \bibinfo {author} {\bibfnamefont {L.-H.}\ \bibnamefont {Lim}}, \ and\ \bibinfo {author} {\bibfnamefont {B.}~\bibnamefont {Mourrain}},\ }\href {\doibase 10.1137/060661569} {\bibfield  {journal} {\bibinfo  {journal} {SIAM Journal on Matrix Analysis and Applications}\ }\textbf {\bibinfo {volume} {30}},\ \bibinfo {pages} {1254} (\bibinfo {year} {2008})}\BibitemShut {NoStop}%
\bibitem [{\citenamefont {Shimony}(1995)}]{shimony1995degree}%
  \BibitemOpen
  \bibfield  {author} {\bibinfo {author} {\bibfnamefont {A.}~\bibnamefont {Shimony}},\ }\href {\doibase 10.1111/j.1749-6632.1995.tb39008.x} {\bibfield  {journal} {\bibinfo  {journal} {Annals of the New York Academy of Sciences}\ }\textbf {\bibinfo {volume} {755}},\ \bibinfo {pages} {675} (\bibinfo {year} {1995})}\BibitemShut {NoStop}%
\bibitem [{\citenamefont {Wei}\ and\ \citenamefont {Goldbart}(2003)}]{wei2003geometric}%
  \BibitemOpen
  \bibfield  {author} {\bibinfo {author} {\bibfnamefont {T.-C.}\ \bibnamefont {Wei}}\ and\ \bibinfo {author} {\bibfnamefont {P.~M.}\ \bibnamefont {Goldbart}},\ }\href {\doibase 10.1103/PhysRevA.68.042307} {\bibfield  {journal} {\bibinfo  {journal} {Physical Review A}\ }\textbf {\bibinfo {volume} {68}},\ \bibinfo {pages} {042307} (\bibinfo {year} {2003})}\BibitemShut {NoStop}%
\bibitem [{\citenamefont {Weinbrenner}\ and\ \citenamefont {G{\"u}hne}(2025)}]{weinbrenner2025quantifying}%
  \BibitemOpen
  \bibfield  {author} {\bibinfo {author} {\bibfnamefont {L.~T.}\ \bibnamefont {Weinbrenner}}\ and\ \bibinfo {author} {\bibfnamefont {O.}~\bibnamefont {G{\"u}hne}},\ }\href {\doibase 10.1209/0295-5075/adffb5} {\bibfield  {journal} {\bibinfo  {journal} {Europhysics Letters}\ }\textbf {\bibinfo {volume} {151}},\ \bibinfo {pages} {68001} (\bibinfo {year} {2025})}\BibitemShut {NoStop}%
\bibitem [{\citenamefont {Hayashi}\ \emph {et~al.}(2006)\citenamefont {Hayashi}, \citenamefont {Markham}, \citenamefont {Murao}, \citenamefont {Owari},\ and\ \citenamefont {Virmani}}]{hayashi2006bounds}%
  \BibitemOpen
  \bibfield  {author} {\bibinfo {author} {\bibfnamefont {M.}~\bibnamefont {Hayashi}}, \bibinfo {author} {\bibfnamefont {D.}~\bibnamefont {Markham}}, \bibinfo {author} {\bibfnamefont {M.}~\bibnamefont {Murao}}, \bibinfo {author} {\bibfnamefont {M.}~\bibnamefont {Owari}}, \ and\ \bibinfo {author} {\bibfnamefont {S.}~\bibnamefont {Virmani}},\ }\href {\doibase 10.1103/PhysRevLett.96.040501} {\bibfield  {journal} {\bibinfo  {journal} {Physical Review Letters}\ }\textbf {\bibinfo {volume} {96}},\ \bibinfo {pages} {040501} (\bibinfo {year} {2006})}\BibitemShut {NoStop}%
\bibitem [{\citenamefont {De~Silva}\ and\ \citenamefont {Lim}(2008)}]{de2008tensor}%
  \BibitemOpen
  \bibfield  {author} {\bibinfo {author} {\bibfnamefont {V.}~\bibnamefont {De~Silva}}\ and\ \bibinfo {author} {\bibfnamefont {L.-H.}\ \bibnamefont {Lim}},\ }\href {\doibase 10.1137/06066518X} {\bibfield  {journal} {\bibinfo  {journal} {SIAM Journal on Matrix Analysis and Applications}\ }\textbf {\bibinfo {volume} {30}},\ \bibinfo {pages} {1084} (\bibinfo {year} {2008})}\BibitemShut {NoStop}%
\bibitem [{\citenamefont {H{\"u}bener}\ \emph {et~al.}(2009)\citenamefont {H{\"u}bener}, \citenamefont {Kleinmann}, \citenamefont {Wei}, \citenamefont {Gonz{\'a}lez-Guill{\'e}n},\ and\ \citenamefont {G{\"u}hne}}]{hubener2009geometric}%
  \BibitemOpen
  \bibfield  {author} {\bibinfo {author} {\bibfnamefont {R.}~\bibnamefont {H{\"u}bener}}, \bibinfo {author} {\bibfnamefont {M.}~\bibnamefont {Kleinmann}}, \bibinfo {author} {\bibfnamefont {T.-C.}\ \bibnamefont {Wei}}, \bibinfo {author} {\bibfnamefont {C.}~\bibnamefont {Gonz{\'a}lez-Guill{\'e}n}}, \ and\ \bibinfo {author} {\bibfnamefont {O.}~\bibnamefont {G{\"u}hne}},\ }\href {\doibase 10.1103/PhysRevA.80.032324} {\bibfield  {journal} {\bibinfo  {journal} {Physical Review A—Atomic, Molecular, and Optical Physics}\ }\textbf {\bibinfo {volume} {80}},\ \bibinfo {pages} {032324} (\bibinfo {year} {2009})}\BibitemShut {NoStop}%
\bibitem [{\citenamefont {Giraud}\ \emph {et~al.}(2015)\citenamefont {Giraud}, \citenamefont {Braun}, \citenamefont {Baguette}, \citenamefont {Bastin},\ and\ \citenamefont {Martin}}]{giraud2015tensor}%
  \BibitemOpen
  \bibfield  {author} {\bibinfo {author} {\bibfnamefont {O.}~\bibnamefont {Giraud}}, \bibinfo {author} {\bibfnamefont {D.}~\bibnamefont {Braun}}, \bibinfo {author} {\bibfnamefont {D.}~\bibnamefont {Baguette}}, \bibinfo {author} {\bibfnamefont {T.}~\bibnamefont {Bastin}}, \ and\ \bibinfo {author} {\bibfnamefont {J.}~\bibnamefont {Martin}},\ }\href {\doibase 10.1103/PhysRevLett.114.080401} {\bibfield  {journal} {\bibinfo  {journal} {Physical Review Letters}\ }\textbf {\bibinfo {volume} {114}},\ \bibinfo {pages} {080401} (\bibinfo {year} {2015})}\BibitemShut {NoStop}%
\bibitem [{\citenamefont {Liu}\ \emph {et~al.}(2012)\citenamefont {Liu}, \citenamefont {Li}, \citenamefont {Li},\ and\ \citenamefont {Qiao}}]{liu2012local}%
  \BibitemOpen
  \bibfield  {author} {\bibinfo {author} {\bibfnamefont {B.}~\bibnamefont {Liu}}, \bibinfo {author} {\bibfnamefont {J.-L.}\ \bibnamefont {Li}}, \bibinfo {author} {\bibfnamefont {X.}~\bibnamefont {Li}}, \ and\ \bibinfo {author} {\bibfnamefont {C.-F.}\ \bibnamefont {Qiao}},\ }\href {\doibase 10.1103/PhysRevLett.108.050501} {\bibfield  {journal} {\bibinfo  {journal} {Physical Review Letters}\ }\textbf {\bibinfo {volume} {108}},\ \bibinfo {pages} {050501} (\bibinfo {year} {2012})}\BibitemShut {NoStop}%
\bibitem [{\citenamefont {Bastin}\ \emph {et~al.}(2009)\citenamefont {Bastin}, \citenamefont {Krins}, \citenamefont {Mathonet}, \citenamefont {Godefroid}, \citenamefont {Lamata},\ and\ \citenamefont {Solano}}]{bastin2009operational}%
  \BibitemOpen
  \bibfield  {author} {\bibinfo {author} {\bibfnamefont {T.}~\bibnamefont {Bastin}}, \bibinfo {author} {\bibfnamefont {S.}~\bibnamefont {Krins}}, \bibinfo {author} {\bibfnamefont {P.}~\bibnamefont {Mathonet}}, \bibinfo {author} {\bibfnamefont {M.}~\bibnamefont {Godefroid}}, \bibinfo {author} {\bibfnamefont {L.}~\bibnamefont {Lamata}}, \ and\ \bibinfo {author} {\bibfnamefont {E.}~\bibnamefont {Solano}},\ }\href {\doibase 10.1103/PhysRevLett.103.070503} {\bibfield  {journal} {\bibinfo  {journal} {Physical Review Letters}\ }\textbf {\bibinfo {volume} {103}},\ \bibinfo {pages} {070503} (\bibinfo {year} {2009})}\BibitemShut {NoStop}%
\bibitem [{\citenamefont {Aulbach}\ \emph {et~al.}(2010)\citenamefont {Aulbach}, \citenamefont {Markham},\ and\ \citenamefont {Murao}}]{aulbach2010maximally}%
  \BibitemOpen
  \bibfield  {author} {\bibinfo {author} {\bibfnamefont {M.}~\bibnamefont {Aulbach}}, \bibinfo {author} {\bibfnamefont {D.}~\bibnamefont {Markham}}, \ and\ \bibinfo {author} {\bibfnamefont {M.}~\bibnamefont {Murao}},\ }\href {\doibase 10.1088/1367-2630/12/7/073025} {\bibfield  {journal} {\bibinfo  {journal} {New Journal of Physics}\ }\textbf {\bibinfo {volume} {12}},\ \bibinfo {pages} {073025} (\bibinfo {year} {2010})}\BibitemShut {NoStop}%
\bibitem [{\citenamefont {Martin}\ \emph {et~al.}(2010)\citenamefont {Martin}, \citenamefont {Giraud}, \citenamefont {Braun}, \citenamefont {Braun},\ and\ \citenamefont {Bastin}}]{martin2010multiqubit}%
  \BibitemOpen
  \bibfield  {author} {\bibinfo {author} {\bibfnamefont {J.}~\bibnamefont {Martin}}, \bibinfo {author} {\bibfnamefont {O.}~\bibnamefont {Giraud}}, \bibinfo {author} {\bibfnamefont {P.}~\bibnamefont {Braun}}, \bibinfo {author} {\bibfnamefont {D.}~\bibnamefont {Braun}}, \ and\ \bibinfo {author} {\bibfnamefont {T.}~\bibnamefont {Bastin}},\ }\href {\doibase 10.1103/PhysRevA.81.062347} {\bibfield  {journal} {\bibinfo  {journal} {Physical Review A—Atomic, Molecular, and Optical Physics}\ }\textbf {\bibinfo {volume} {81}},\ \bibinfo {pages} {062347} (\bibinfo {year} {2010})}\BibitemShut {NoStop}%
\bibitem [{\citenamefont {Markham}(2011)}]{markham2011entanglement}%
  \BibitemOpen
  \bibfield  {author} {\bibinfo {author} {\bibfnamefont {D.~J.}\ \bibnamefont {Markham}},\ }\href {\doibase 10.1103/PhysRevA.83.042332} {\bibfield  {journal} {\bibinfo  {journal} {Physical Review A—Atomic, Molecular, and Optical Physics}\ }\textbf {\bibinfo {volume} {83}},\ \bibinfo {pages} {042332} (\bibinfo {year} {2011})}\BibitemShut {NoStop}%
\bibitem [{\citenamefont {Ribeiro}\ and\ \citenamefont {Mosseri}(2011)}]{ribeiro2011entanglement}%
  \BibitemOpen
  \bibfield  {author} {\bibinfo {author} {\bibfnamefont {P.}~\bibnamefont {Ribeiro}}\ and\ \bibinfo {author} {\bibfnamefont {R.}~\bibnamefont {Mosseri}},\ }\href {\doibase 10.1103/PhysRevLett.106.180502} {\bibfield  {journal} {\bibinfo  {journal} {Physical Review Letters}\ }\textbf {\bibinfo {volume} {106}},\ \bibinfo {pages} {180502} (\bibinfo {year} {2011})}\BibitemShut {NoStop}%
\bibitem [{\citenamefont {Wang}\ and\ \citenamefont {Markham}(2012)}]{wang2012nonlocality}%
  \BibitemOpen
  \bibfield  {author} {\bibinfo {author} {\bibfnamefont {Z.}~\bibnamefont {Wang}}\ and\ \bibinfo {author} {\bibfnamefont {D.}~\bibnamefont {Markham}},\ }\href {\doibase 10.1103/PhysRevLett.108.210407} {\bibfield  {journal} {\bibinfo  {journal} {Physical Review Letters}\ }\textbf {\bibinfo {volume} {108}},\ \bibinfo {pages} {210407} (\bibinfo {year} {2012})}\BibitemShut {NoStop}%
\bibitem [{\citenamefont {Ohst}\ \emph {et~al.}(2025)\citenamefont {Ohst}, \citenamefont {Yadin}, \citenamefont {Ostermann}, \citenamefont {de~Wolff}, \citenamefont {G{\"u}hne},\ and\ \citenamefont {Nguyen}}]{ohst2025revealing}%
  \BibitemOpen
  \bibfield  {author} {\bibinfo {author} {\bibfnamefont {T.-A.}\ \bibnamefont {Ohst}}, \bibinfo {author} {\bibfnamefont {B.}~\bibnamefont {Yadin}}, \bibinfo {author} {\bibfnamefont {B.}~\bibnamefont {Ostermann}}, \bibinfo {author} {\bibfnamefont {T.}~\bibnamefont {de~Wolff}}, \bibinfo {author} {\bibfnamefont {O.}~\bibnamefont {G{\"u}hne}}, \ and\ \bibinfo {author} {\bibfnamefont {H.-C.}\ \bibnamefont {Nguyen}},\ }\href {\doibase 10.1103/PhysRevLett.134.030201} {\bibfield  {journal} {\bibinfo  {journal} {Physical Review Letters}\ }\textbf {\bibinfo {volume} {134}},\ \bibinfo {pages} {030201} (\bibinfo {year} {2025})}\BibitemShut {NoStop}%
\bibitem [{\citenamefont {Arnaud}\ and\ \citenamefont {Cerf}(2013)}]{arnaud2013exploring}%
  \BibitemOpen
  \bibfield  {author} {\bibinfo {author} {\bibfnamefont {L.}~\bibnamefont {Arnaud}}\ and\ \bibinfo {author} {\bibfnamefont {N.~J.}\ \bibnamefont {Cerf}},\ }\href {\doibase 10.1103/PhysRevA.87.012319} {\bibfield  {journal} {\bibinfo  {journal} {Physical Review A—Atomic, Molecular, and Optical Physics}\ }\textbf {\bibinfo {volume} {87}},\ \bibinfo {pages} {012319} (\bibinfo {year} {2013})}\BibitemShut {NoStop}%
\bibitem [{\citenamefont {Baguette}\ \emph {et~al.}(2014)\citenamefont {Baguette}, \citenamefont {Bastin},\ and\ \citenamefont {Martin}}]{baguette2014multiqubit}%
  \BibitemOpen
  \bibfield  {author} {\bibinfo {author} {\bibfnamefont {D.}~\bibnamefont {Baguette}}, \bibinfo {author} {\bibfnamefont {T.}~\bibnamefont {Bastin}}, \ and\ \bibinfo {author} {\bibfnamefont {J.}~\bibnamefont {Martin}},\ }\href {\doibase 10.1103/PhysRevA.90.032314} {\bibfield  {journal} {\bibinfo  {journal} {Physical Review A}\ }\textbf {\bibinfo {volume} {90}},\ \bibinfo {pages} {032314} (\bibinfo {year} {2014})}\BibitemShut {NoStop}%
\bibitem [{\citenamefont {Marconi}(2023)}]{marconi2023entanglement}%
  \BibitemOpen
  \bibfield  {author} {\bibinfo {author} {\bibfnamefont {C.}~\bibnamefont {Marconi}},\ }\href {http://hdl.handle.net/10803/688705} {\bibfield  {journal} {\bibinfo  {journal} {PhD thesis, Universitat Aut{\`o}noma de Barcelona}\ } (\bibinfo {year} {2023})}\BibitemShut {NoStop}%
\bibitem [{\citenamefont {Romero-Pallejà}\ \emph {et~al.}(2025)\citenamefont {Romero-Pallejà}, \citenamefont {Ahiable}, \citenamefont {Romancino}, \citenamefont {Marconi},\ and\ \citenamefont {Sanpera}}]{romero2024multipartite}%
  \BibitemOpen
  \bibfield  {author} {\bibinfo {author} {\bibfnamefont {J.}~\bibnamefont {Romero-Pallejà}}, \bibinfo {author} {\bibfnamefont {J.}~\bibnamefont {Ahiable}}, \bibinfo {author} {\bibfnamefont {A.}~\bibnamefont {Romancino}}, \bibinfo {author} {\bibfnamefont {C.}~\bibnamefont {Marconi}}, \ and\ \bibinfo {author} {\bibfnamefont {A.}~\bibnamefont {Sanpera}},\ }\href {\doibase 10.1063/5.0240964} {\bibfield  {journal} {\bibinfo  {journal} {Journal of Mathematical Physics}\ }\textbf {\bibinfo {volume} {66}},\ \bibinfo {pages} {022203} (\bibinfo {year} {2025})}\BibitemShut {NoStop}%
\bibitem [{\citenamefont {Marconi}\ and\ \citenamefont {Imai}()}]{inpreparation}%
  \BibitemOpen
  \bibfield  {author} {\bibinfo {author} {\bibfnamefont {C.}~\bibnamefont {Marconi}}\ and\ \bibinfo {author} {\bibfnamefont {S.}~\bibnamefont {Imai}},\ }\href@noop {} {\bibinfo  {journal} {in preparation}\ }\BibitemShut {NoStop}%
\bibitem [{\citenamefont {Eckert}\ \emph {et~al.}(2002)\citenamefont {Eckert}, \citenamefont {Schliemann}, \citenamefont {Bruß},\ and\ \citenamefont {Lewenstein}}]{eckert2002quantum}%
  \BibitemOpen
\bibfield  {journal} {  }\bibfield  {author} {\bibinfo {author} {\bibfnamefont {K.}~\bibnamefont {Eckert}}, \bibinfo {author} {\bibfnamefont {J.}~\bibnamefont {Schliemann}}, \bibinfo {author} {\bibfnamefont {D.}~\bibnamefont {Bruß}}, \ and\ \bibinfo {author} {\bibfnamefont {M.}~\bibnamefont {Lewenstein}},\ }\href {\doibase https://doi.org/10.1006/aphy.2002.6268} {\bibfield  {journal} {\bibinfo  {journal} {Annals of Physics}\ }\textbf {\bibinfo {volume} {299}},\ \bibinfo {pages} {88} (\bibinfo {year} {2002})}\BibitemShut {NoStop}%
\bibitem [{\citenamefont {Ichikawa}\ \emph {et~al.}(2008)\citenamefont {Ichikawa}, \citenamefont {Sasaki}, \citenamefont {Tsutsui},\ and\ \citenamefont {Yonezawa}}]{ichikawa2008exchange}%
  \BibitemOpen
  \bibfield  {author} {\bibinfo {author} {\bibfnamefont {T.}~\bibnamefont {Ichikawa}}, \bibinfo {author} {\bibfnamefont {T.}~\bibnamefont {Sasaki}}, \bibinfo {author} {\bibfnamefont {I.}~\bibnamefont {Tsutsui}}, \ and\ \bibinfo {author} {\bibfnamefont {N.}~\bibnamefont {Yonezawa}},\ }\href {\doibase 10.1103/PhysRevA.78.052105} {\bibfield  {journal} {\bibinfo  {journal} {Physical Review A—Atomic, Molecular, and Optical Physics}\ }\textbf {\bibinfo {volume} {78}},\ \bibinfo {pages} {052105} (\bibinfo {year} {2008})}\BibitemShut {NoStop}%
\bibitem [{\citenamefont {Wei}(2010)}]{wei2010exchange}%
  \BibitemOpen
  \bibfield  {author} {\bibinfo {author} {\bibfnamefont {T.-C.}\ \bibnamefont {Wei}},\ }\href {\doibase 10.1103/PhysRevA.81.054102} {\bibfield  {journal} {\bibinfo  {journal} {Physical Review A—Atomic, Molecular, and Optical Physics}\ }\textbf {\bibinfo {volume} {81}},\ \bibinfo {pages} {054102} (\bibinfo {year} {2010})}\BibitemShut {NoStop}%
\bibitem [{\citenamefont {Peres}(1996)}]{peres1996separability}%
  \BibitemOpen
  \bibfield  {author} {\bibinfo {author} {\bibfnamefont {A.}~\bibnamefont {Peres}},\ }\href {\doibase 10.1103/PhysRevLett.77.1413} {\bibfield  {journal} {\bibinfo  {journal} {Physical Review Letters}\ }\textbf {\bibinfo {volume} {77}},\ \bibinfo {pages} {1413} (\bibinfo {year} {1996})}\BibitemShut {NoStop}%
\bibitem [{\citenamefont {Devi}\ \emph {et~al.}(2007)\citenamefont {Devi}, \citenamefont {Prabhu},\ and\ \citenamefont {Rajagopal}}]{devi2007characterizing}%
  \BibitemOpen
  \bibfield  {author} {\bibinfo {author} {\bibfnamefont {A.~U.}\ \bibnamefont {Devi}}, \bibinfo {author} {\bibfnamefont {R.}~\bibnamefont {Prabhu}}, \ and\ \bibinfo {author} {\bibfnamefont {A.}~\bibnamefont {Rajagopal}},\ }\href {\doibase 10.1103/PhysRevLett.98.060501} {\bibfield  {journal} {\bibinfo  {journal} {Physical Review Letters}\ }\textbf {\bibinfo {volume} {98}},\ \bibinfo {pages} {060501} (\bibinfo {year} {2007})}\BibitemShut {NoStop}%
\bibitem [{\citenamefont {Tura}\ \emph {et~al.}(2018)\citenamefont {Tura}, \citenamefont {Aloy}, \citenamefont {Quesada}, \citenamefont {Lewenstein},\ and\ \citenamefont {Sanpera}}]{tura2018separability}%
  \BibitemOpen
  \bibfield  {author} {\bibinfo {author} {\bibfnamefont {J.}~\bibnamefont {Tura}}, \bibinfo {author} {\bibfnamefont {A.}~\bibnamefont {Aloy}}, \bibinfo {author} {\bibfnamefont {R.}~\bibnamefont {Quesada}}, \bibinfo {author} {\bibfnamefont {M.}~\bibnamefont {Lewenstein}}, \ and\ \bibinfo {author} {\bibfnamefont {A.}~\bibnamefont {Sanpera}},\ }\href {\doibase https://doi.org/10.22331/q-2018-01-12-45} {\bibfield  {journal} {\bibinfo  {journal} {Quantum}\ }\textbf {\bibinfo {volume} {2}},\ \bibinfo {pages} {45} (\bibinfo {year} {2018})}\BibitemShut {NoStop}%
\bibitem [{\citenamefont {Peres}(1997)}]{peres1997quantum}%
  \BibitemOpen
  \bibfield  {author} {\bibinfo {author} {\bibfnamefont {A.}~\bibnamefont {Peres}},\ }\href {\doibase 10.1007/0-306-47120-5} {\emph {\bibinfo {title} {Quantum theory: concepts and methods}}},\ Vol.~\bibinfo {volume} {72}\ (\bibinfo  {publisher} {Springer},\ \bibinfo {year} {1997})\BibitemShut {NoStop}%
\bibitem [{\citenamefont {Nielsen}\ and\ \citenamefont {Chuang}(2010)}]{nielsen_chuang_2010}%
  \BibitemOpen
  \bibfield  {author} {\bibinfo {author} {\bibfnamefont {M.~A.}\ \bibnamefont {Nielsen}}\ and\ \bibinfo {author} {\bibfnamefont {I.~L.}\ \bibnamefont {Chuang}},\ }\href {\doibase 10.1017/CBO9780511976667} {\emph {\bibinfo {title} {Quantum Computation and Quantum Information: 10th Anniversary Edition}}}\ (\bibinfo  {publisher} {Cambridge University Press},\ \bibinfo {year} {2010})\BibitemShut {NoStop}%
\bibitem [{\citenamefont {Bini}\ \emph {et~al.}(1980)\citenamefont {Bini}, \citenamefont {Lotti},\ and\ \citenamefont {Romani}}]{bini1980approximate}%
  \BibitemOpen
  \bibfield  {author} {\bibinfo {author} {\bibfnamefont {D.}~\bibnamefont {Bini}}, \bibinfo {author} {\bibfnamefont {G.}~\bibnamefont {Lotti}}, \ and\ \bibinfo {author} {\bibfnamefont {F.}~\bibnamefont {Romani}},\ }\href {\doibase 10.1137/0209053} {\bibfield  {journal} {\bibinfo  {journal} {SIAM Journal on Computing}\ }\textbf {\bibinfo {volume} {9}},\ \bibinfo {pages} {692} (\bibinfo {year} {1980})}\BibitemShut {NoStop}%
\bibitem [{\citenamefont {Landsberg}(2011)}]{landsberg2011tensors}%
  \BibitemOpen
  \bibfield  {author} {\bibinfo {author} {\bibfnamefont {J.~M.}\ \bibnamefont {Landsberg}},\ }\href {\doibase 10.1090/gsm/128} {\emph {\bibinfo {title} {Tensors: geometry and applications: geometry and applications}}},\ Vol.\ \bibinfo {volume} {128}\ (\bibinfo  {publisher} {American Mathematical Soc.},\ \bibinfo {year} {2011})\BibitemShut {NoStop}%
\bibitem [{\citenamefont {Allman}\ \emph {et~al.}(2013)\citenamefont {Allman}, \citenamefont {Jarvis}, \citenamefont {Rhodes},\ and\ \citenamefont {Sumner}}]{allman2013tensor}%
  \BibitemOpen
  \bibfield  {author} {\bibinfo {author} {\bibfnamefont {E.~S.}\ \bibnamefont {Allman}}, \bibinfo {author} {\bibfnamefont {P.~D.}\ \bibnamefont {Jarvis}}, \bibinfo {author} {\bibfnamefont {J.~A.}\ \bibnamefont {Rhodes}}, \ and\ \bibinfo {author} {\bibfnamefont {J.~G.}\ \bibnamefont {Sumner}},\ }\href {\doibase 10.1137/120899066} {\bibfield  {journal} {\bibinfo  {journal} {SIAM Journal on Matrix Analysis and Applications}\ }\textbf {\bibinfo {volume} {34}},\ \bibinfo {pages} {1014} (\bibinfo {year} {2013})}\BibitemShut {NoStop}%
\bibitem [{\citenamefont {Vrana}\ and\ \citenamefont {Christandl}(2015)}]{vrana2015asymptotic}%
  \BibitemOpen
  \bibfield  {author} {\bibinfo {author} {\bibfnamefont {P.}~\bibnamefont {Vrana}}\ and\ \bibinfo {author} {\bibfnamefont {M.}~\bibnamefont {Christandl}},\ }\href {\doibase 10.1063/1.4908106} {\bibfield  {journal} {\bibinfo  {journal} {Journal of Mathematical Physics}\ }\textbf {\bibinfo {volume} {56}} (\bibinfo {year} {2015}),\ 10.1063/1.4908106}\BibitemShut {NoStop}%
\bibitem [{\citenamefont {Zuiddam}(2017)}]{zuiddam2017note}%
  \BibitemOpen
  \bibfield  {author} {\bibinfo {author} {\bibfnamefont {J.}~\bibnamefont {Zuiddam}},\ }\href {\doibase 10.1016/j.laa.2017.03.015} {\bibfield  {journal} {\bibinfo  {journal} {Linear Algebra and its Applications}\ }\textbf {\bibinfo {volume} {525}},\ \bibinfo {pages} {33} (\bibinfo {year} {2017})}\BibitemShut {NoStop}%
\bibitem [{\citenamefont {Bruzda}\ \emph {et~al.}(2024)\citenamefont {Bruzda}, \citenamefont {Friedland},\ and\ \citenamefont {{\.Z}yczkowski}}]{bruzda2024rank}%
  \BibitemOpen
  \bibfield  {author} {\bibinfo {author} {\bibfnamefont {W.}~\bibnamefont {Bruzda}}, \bibinfo {author} {\bibfnamefont {S.}~\bibnamefont {Friedland}}, \ and\ \bibinfo {author} {\bibfnamefont {K.}~\bibnamefont {{\.Z}yczkowski}},\ }\href {\doibase https://doi.org/10.1080/03081087.2023.2211717} {\bibfield  {journal} {\bibinfo  {journal} {Linear and Multilinear Algebra}\ }\textbf {\bibinfo {volume} {72}},\ \bibinfo {pages} {1796} (\bibinfo {year} {2024})}\BibitemShut {NoStop}%
\bibitem [{\citenamefont {Landsberg}\ and\ \citenamefont {Teitler}(2010)}]{landsberg2010ranks}%
  \BibitemOpen
  \bibfield  {author} {\bibinfo {author} {\bibfnamefont {J.~M.}\ \bibnamefont {Landsberg}}\ and\ \bibinfo {author} {\bibfnamefont {Z.}~\bibnamefont {Teitler}},\ }\href {\doibase 10.1007/s10208-009-9055-3} {\bibfield  {journal} {\bibinfo  {journal} {Foundations of Computational Mathematics}\ }\textbf {\bibinfo {volume} {10}},\ \bibinfo {pages} {339} (\bibinfo {year} {2010})}\BibitemShut {NoStop}%
\bibitem [{\citenamefont {Buczy{\'n}ski}\ \emph {et~al.}(2013)\citenamefont {Buczy{\'n}ski}, \citenamefont {Ginensky},\ and\ \citenamefont {Landsberg}}]{buczynski2013determinantal}%
  \BibitemOpen
  \bibfield  {author} {\bibinfo {author} {\bibfnamefont {J.}~\bibnamefont {Buczy{\'n}ski}}, \bibinfo {author} {\bibfnamefont {A.}~\bibnamefont {Ginensky}}, \ and\ \bibinfo {author} {\bibfnamefont {J.~M.}\ \bibnamefont {Landsberg}},\ }\href {\doibase https://doi.org/10.1112/jlms/jds073} {\bibfield  {journal} {\bibinfo  {journal} {Journal of the London Mathematical Society}\ }\textbf {\bibinfo {volume} {88}},\ \bibinfo {pages} {1} (\bibinfo {year} {2013})}\BibitemShut {NoStop}%
\bibitem [{\citenamefont {Friedland}(2016)}]{friedland2016remarks}%
  \BibitemOpen
  \bibfield  {author} {\bibinfo {author} {\bibfnamefont {S.}~\bibnamefont {Friedland}},\ }\href {\doibase https://doi.org/10.1137/15M1022653} {\bibfield  {journal} {\bibinfo  {journal} {SIAM Journal on Matrix Analysis and Applications}\ }\textbf {\bibinfo {volume} {37}},\ \bibinfo {pages} {320} (\bibinfo {year} {2016})}\BibitemShut {NoStop}%
\bibitem [{\citenamefont {Zhang}\ \emph {et~al.}(2016)\citenamefont {Zhang}, \citenamefont {Huang},\ and\ \citenamefont {Qi}}]{zhang2016comon}%
  \BibitemOpen
  \bibfield  {author} {\bibinfo {author} {\bibfnamefont {X.}~\bibnamefont {Zhang}}, \bibinfo {author} {\bibfnamefont {Z.-H.}\ \bibnamefont {Huang}}, \ and\ \bibinfo {author} {\bibfnamefont {L.}~\bibnamefont {Qi}},\ }\href {\doibase 10.1137/141001470} {\bibfield  {journal} {\bibinfo  {journal} {SIAM Journal on Matrix Analysis and Applications}\ }\textbf {\bibinfo {volume} {37}},\ \bibinfo {pages} {1719} (\bibinfo {year} {2016})}\BibitemShut {NoStop}%
\bibitem [{\citenamefont {Bourennane}\ \emph {et~al.}(2004)\citenamefont {Bourennane}, \citenamefont {Eibl}, \citenamefont {Kurtsiefer}, \citenamefont {Gaertner}, \citenamefont {Weinfurter}, \citenamefont {G{\"u}hne}, \citenamefont {Hyllus}, \citenamefont {Bru{\ss}}, \citenamefont {Lewenstein},\ and\ \citenamefont {Sanpera}}]{bourennane2004experimental}%
  \BibitemOpen
  \bibfield  {author} {\bibinfo {author} {\bibfnamefont {M.}~\bibnamefont {Bourennane}}, \bibinfo {author} {\bibfnamefont {M.}~\bibnamefont {Eibl}}, \bibinfo {author} {\bibfnamefont {C.}~\bibnamefont {Kurtsiefer}}, \bibinfo {author} {\bibfnamefont {S.}~\bibnamefont {Gaertner}}, \bibinfo {author} {\bibfnamefont {H.}~\bibnamefont {Weinfurter}}, \bibinfo {author} {\bibfnamefont {O.}~\bibnamefont {G{\"u}hne}}, \bibinfo {author} {\bibfnamefont {P.}~\bibnamefont {Hyllus}}, \bibinfo {author} {\bibfnamefont {D.}~\bibnamefont {Bru{\ss}}}, \bibinfo {author} {\bibfnamefont {M.}~\bibnamefont {Lewenstein}}, \ and\ \bibinfo {author} {\bibfnamefont {A.}~\bibnamefont {Sanpera}},\ }\href {\doibase 10.1103/PhysRevLett.92.087902} {\bibfield  {journal} {\bibinfo  {journal} {Physical Review Letters}\ }\textbf {\bibinfo {volume} {92}},\ \bibinfo {pages} {087902} (\bibinfo {year} {2004})}\BibitemShut {NoStop}%
\bibitem [{\citenamefont {Aulbach}(2012)}]{aulbach2012classification}%
  \BibitemOpen
  \bibfield  {author} {\bibinfo {author} {\bibfnamefont {M.}~\bibnamefont {Aulbach}},\ }\href {\doibase 10.1142/S0219749912300045} {\bibfield  {journal} {\bibinfo  {journal} {International Journal of Quantum Information}\ }\textbf {\bibinfo {volume} {10}},\ \bibinfo {pages} {1230004} (\bibinfo {year} {2012})}\BibitemShut {NoStop}%
\bibitem [{\citenamefont {Derksen}\ \emph {et~al.}(2017)\citenamefont {Derksen}, \citenamefont {Friedland}, \citenamefont {Lim},\ and\ \citenamefont {Wang}}]{derksen2017theoretical}%
  \BibitemOpen
  \bibfield  {author} {\bibinfo {author} {\bibfnamefont {H.}~\bibnamefont {Derksen}}, \bibinfo {author} {\bibfnamefont {S.}~\bibnamefont {Friedland}}, \bibinfo {author} {\bibfnamefont {L.-H.}\ \bibnamefont {Lim}}, \ and\ \bibinfo {author} {\bibfnamefont {L.}~\bibnamefont {Wang}},\ }\href {https://arxiv.org/abs/1705.07160} {\bibfield  {journal} {\bibinfo  {journal} {arXiv preprint arXiv:1705.07160}\ } (\bibinfo {year} {2017})}\BibitemShut {NoStop}%
\bibitem [{\citenamefont {Friedland}\ and\ \citenamefont {Lim}(2018)}]{friedland2018nuclear}%
  \BibitemOpen
  \bibfield  {author} {\bibinfo {author} {\bibfnamefont {S.}~\bibnamefont {Friedland}}\ and\ \bibinfo {author} {\bibfnamefont {L.-H.}\ \bibnamefont {Lim}},\ }\href {\doibase https://doi.org/10.1090/mcom/3239} {\bibfield  {journal} {\bibinfo  {journal} {Mathematics of Computation}\ }\textbf {\bibinfo {volume} {87}},\ \bibinfo {pages} {1255} (\bibinfo {year} {2018})}\BibitemShut {NoStop}%
\bibitem [{\citenamefont {Lim}\ and\ \citenamefont {Comon}(2013)}]{lim2013blind}%
  \BibitemOpen
  \bibfield  {author} {\bibinfo {author} {\bibfnamefont {L.-H.}\ \bibnamefont {Lim}}\ and\ \bibinfo {author} {\bibfnamefont {P.}~\bibnamefont {Comon}},\ }\href {\doibase 10.1109/TIT.2013.2291876} {\bibfield  {journal} {\bibinfo  {journal} {IEEE Transactions on Information Theory}\ }\textbf {\bibinfo {volume} {60}},\ \bibinfo {pages} {1260} (\bibinfo {year} {2013})}\BibitemShut {NoStop}%
\bibitem [{\citenamefont {Vidal}\ and\ \citenamefont {Werner}(2002)}]{vidal2002computable}%
  \BibitemOpen
  \bibfield  {author} {\bibinfo {author} {\bibfnamefont {G.}~\bibnamefont {Vidal}}\ and\ \bibinfo {author} {\bibfnamefont {R.~F.}\ \bibnamefont {Werner}},\ }\href {\doibase https://doi.org/10.1103/PhysRevA.65.032314} {\bibfield  {journal} {\bibinfo  {journal} {Physical Review A}\ }\textbf {\bibinfo {volume} {65}},\ \bibinfo {pages} {032314} (\bibinfo {year} {2002})}\BibitemShut {NoStop}%
\bibitem [{\citenamefont {Parthasarathy}(2004)}]{parthasarathy2004maximal}%
  \BibitemOpen
  \bibfield  {author} {\bibinfo {author} {\bibfnamefont {K.~R.}\ \bibnamefont {Parthasarathy}},\ }\href {\doibase 10.1007/BF02829441} {\bibfield  {journal} {\bibinfo  {journal} {Proceedings Mathematical Sciences}\ }\textbf {\bibinfo {volume} {114}},\ \bibinfo {pages} {365} (\bibinfo {year} {2004})}\BibitemShut {NoStop}%
\bibitem [{\citenamefont {Bhat}(2006)}]{bhat2006completely}%
  \BibitemOpen
  \bibfield  {author} {\bibinfo {author} {\bibfnamefont {B.~R.}\ \bibnamefont {Bhat}},\ }\href {\doibase 10.1142/S0219749906001797} {\bibfield  {journal} {\bibinfo  {journal} {International Journal of Quantum Information}\ }\textbf {\bibinfo {volume} {4}},\ \bibinfo {pages} {325} (\bibinfo {year} {2006})}\BibitemShut {NoStop}%
\bibitem [{\citenamefont {Cubitt}\ \emph {et~al.}(2008)\citenamefont {Cubitt}, \citenamefont {Montanaro},\ and\ \citenamefont {Winter}}]{cubitt2008dimension}%
  \BibitemOpen
  \bibfield  {author} {\bibinfo {author} {\bibfnamefont {T.}~\bibnamefont {Cubitt}}, \bibinfo {author} {\bibfnamefont {A.}~\bibnamefont {Montanaro}}, \ and\ \bibinfo {author} {\bibfnamefont {A.}~\bibnamefont {Winter}},\ }\href {\doibase 10.1063/1.2862998} {\bibfield  {journal} {\bibinfo  {journal} {Journal of Mathematical Physics}\ }\textbf {\bibinfo {volume} {49}} (\bibinfo {year} {2008}),\ 10.1063/1.2862998}\BibitemShut {NoStop}%
\bibitem [{\citenamefont {Kunkel}\ \emph {et~al.}(2018)\citenamefont {Kunkel}, \citenamefont {Pr{\"u}fer}, \citenamefont {Strobel}, \citenamefont {Linnemann}, \citenamefont {Fr{\"o}lian}, \citenamefont {Gasenzer}, \citenamefont {G{\"a}rttner},\ and\ \citenamefont {Oberthaler}}]{kunkel2018spatially}%
  \BibitemOpen
  \bibfield  {author} {\bibinfo {author} {\bibfnamefont {P.}~\bibnamefont {Kunkel}}, \bibinfo {author} {\bibfnamefont {M.}~\bibnamefont {Pr{\"u}fer}}, \bibinfo {author} {\bibfnamefont {H.}~\bibnamefont {Strobel}}, \bibinfo {author} {\bibfnamefont {D.}~\bibnamefont {Linnemann}}, \bibinfo {author} {\bibfnamefont {A.}~\bibnamefont {Fr{\"o}lian}}, \bibinfo {author} {\bibfnamefont {T.}~\bibnamefont {Gasenzer}}, \bibinfo {author} {\bibfnamefont {M.}~\bibnamefont {G{\"a}rttner}}, \ and\ \bibinfo {author} {\bibfnamefont {M.~K.}\ \bibnamefont {Oberthaler}},\ }\href {\doibase 10.1126/science.aao2254} {\bibfield  {journal} {\bibinfo  {journal} {Science}\ }\textbf {\bibinfo {volume} {360}},\ \bibinfo {pages} {413} (\bibinfo {year} {2018})}\BibitemShut {NoStop}%
\bibitem [{\citenamefont {Fadel}\ \emph {et~al.}(2018)\citenamefont {Fadel}, \citenamefont {Zibold}, \citenamefont {D{\'e}camps},\ and\ \citenamefont {Treutlein}}]{fadel2018spatial}%
  \BibitemOpen
  \bibfield  {author} {\bibinfo {author} {\bibfnamefont {M.}~\bibnamefont {Fadel}}, \bibinfo {author} {\bibfnamefont {T.}~\bibnamefont {Zibold}}, \bibinfo {author} {\bibfnamefont {B.}~\bibnamefont {D{\'e}camps}}, \ and\ \bibinfo {author} {\bibfnamefont {P.}~\bibnamefont {Treutlein}},\ }\href {\doibase 10.1126/science.aao185} {\bibfield  {journal} {\bibinfo  {journal} {Science}\ }\textbf {\bibinfo {volume} {360}},\ \bibinfo {pages} {409} (\bibinfo {year} {2018})}\BibitemShut {NoStop}%
\bibitem [{\citenamefont {Lange}\ \emph {et~al.}(2018)\citenamefont {Lange}, \citenamefont {Peise}, \citenamefont {L{\"u}cke}, \citenamefont {Kruse}, \citenamefont {Vitagliano}, \citenamefont {Apellaniz}, \citenamefont {Kleinmann}, \citenamefont {T{\'o}th},\ and\ \citenamefont {Klempt}}]{lange2018entanglement}%
  \BibitemOpen
  \bibfield  {author} {\bibinfo {author} {\bibfnamefont {K.}~\bibnamefont {Lange}}, \bibinfo {author} {\bibfnamefont {J.}~\bibnamefont {Peise}}, \bibinfo {author} {\bibfnamefont {B.}~\bibnamefont {L{\"u}cke}}, \bibinfo {author} {\bibfnamefont {I.}~\bibnamefont {Kruse}}, \bibinfo {author} {\bibfnamefont {G.}~\bibnamefont {Vitagliano}}, \bibinfo {author} {\bibfnamefont {I.}~\bibnamefont {Apellaniz}}, \bibinfo {author} {\bibfnamefont {M.}~\bibnamefont {Kleinmann}}, \bibinfo {author} {\bibfnamefont {G.}~\bibnamefont {T{\'o}th}}, \ and\ \bibinfo {author} {\bibfnamefont {C.}~\bibnamefont {Klempt}},\ }\href {\doibase 10.1126/science.aao2035} {\bibfield  {journal} {\bibinfo  {journal} {Science}\ }\textbf {\bibinfo {volume} {360}},\ \bibinfo {pages} {416} (\bibinfo {year} {2018})}\BibitemShut {NoStop}%
\bibitem [{\citenamefont {Stockton}\ \emph {et~al.}(2003)\citenamefont {Stockton}, \citenamefont {Geremia}, \citenamefont {Doherty},\ and\ \citenamefont {Mabuchi}}]{stockton2003characterizing}%
  \BibitemOpen
  \bibfield  {author} {\bibinfo {author} {\bibfnamefont {J.~K.}\ \bibnamefont {Stockton}}, \bibinfo {author} {\bibfnamefont {J.~M.}\ \bibnamefont {Geremia}}, \bibinfo {author} {\bibfnamefont {A.~C.}\ \bibnamefont {Doherty}}, \ and\ \bibinfo {author} {\bibfnamefont {H.}~\bibnamefont {Mabuchi}},\ }\href {\doibase 10.1103/PhysRevA.67.022112} {\bibfield  {journal} {\bibinfo  {journal} {Physical Review A}\ }\textbf {\bibinfo {volume} {67}},\ \bibinfo {pages} {022112} (\bibinfo {year} {2003})}\BibitemShut {NoStop}%
\bibitem [{\citenamefont {T{\'o}th}(2007)}]{toth2007detection}%
  \BibitemOpen
  \bibfield  {author} {\bibinfo {author} {\bibfnamefont {G.}~\bibnamefont {T{\'o}th}},\ }\href {\doibase 10.1364/JOSAB.24.000275} {\bibfield  {journal} {\bibinfo  {journal} {Journal of the Optical Society of America B}\ }\textbf {\bibinfo {volume} {24}},\ \bibinfo {pages} {275} (\bibinfo {year} {2007})}\BibitemShut {NoStop}%
\bibitem [{\citenamefont {Raveh}\ and\ \citenamefont {Nepomechie}(2024)}]{raveh2024dicke}%
  \BibitemOpen
  \bibfield  {author} {\bibinfo {author} {\bibfnamefont {D.}~\bibnamefont {Raveh}}\ and\ \bibinfo {author} {\bibfnamefont {R.~I.}\ \bibnamefont {Nepomechie}},\ }\href {\doibase 10.1103/PhysRevA.110.052438} {\bibfield  {journal} {\bibinfo  {journal} {Physical Review A}\ }\textbf {\bibinfo {volume} {110}},\ \bibinfo {pages} {052438} (\bibinfo {year} {2024})}\BibitemShut {NoStop}%
\bibitem [{\citenamefont {Bernardi}\ and\ \citenamefont {Carusotto}(2012)}]{bernardi2012algebraic}%
  \BibitemOpen
  \bibfield  {author} {\bibinfo {author} {\bibfnamefont {A.}~\bibnamefont {Bernardi}}\ and\ \bibinfo {author} {\bibfnamefont {I.}~\bibnamefont {Carusotto}},\ }\href {\doibase 10.1088/1751-8113/45/10/105304} {\bibfield  {journal} {\bibinfo  {journal} {Journal of Physics A: Mathematical and Theoretical}\ }\textbf {\bibinfo {volume} {45}},\ \bibinfo {pages} {105304} (\bibinfo {year} {2012})}\BibitemShut {NoStop}%
\bibitem [{\citenamefont {Sylvester}(1886)}]{sylvester1886extension}%
  \BibitemOpen
  \bibfield  {author} {\bibinfo {author} {\bibfnamefont {J.~J.}\ \bibnamefont {Sylvester}},\ }\href@noop {} {\bibfield  {journal} {\bibinfo  {journal} {Comptes Rendus, Math. Acad. Sci. Paris}\ }\textbf {\bibinfo {volume} {102}},\ \bibinfo {pages} {1532} (\bibinfo {year} {1886})}\BibitemShut {NoStop}%
\bibitem [{\citenamefont {Brachat}\ \emph {et~al.}(2010)\citenamefont {Brachat}, \citenamefont {Comon}, \citenamefont {Mourrain},\ and\ \citenamefont {Tsigaridas}}]{brachat2010symmetric}%
  \BibitemOpen
  \bibfield  {author} {\bibinfo {author} {\bibfnamefont {J.}~\bibnamefont {Brachat}}, \bibinfo {author} {\bibfnamefont {P.}~\bibnamefont {Comon}}, \bibinfo {author} {\bibfnamefont {B.}~\bibnamefont {Mourrain}}, \ and\ \bibinfo {author} {\bibfnamefont {E.}~\bibnamefont {Tsigaridas}},\ }\href {\doibase 10.1016/j.laa.2010.06.046} {\bibfield  {journal} {\bibinfo  {journal} {Linear Algebra and its Applications}\ }\textbf {\bibinfo {volume} {433}},\ \bibinfo {pages} {1851} (\bibinfo {year} {2010})}\BibitemShut {NoStop}%
\bibitem [{\citenamefont {Comas}\ and\ \citenamefont {Seiguer}(2011)}]{comas2011rank}%
  \BibitemOpen
  \bibfield  {author} {\bibinfo {author} {\bibfnamefont {G.}~\bibnamefont {Comas}}\ and\ \bibinfo {author} {\bibfnamefont {M.}~\bibnamefont {Seiguer}},\ }\href {\doibase 10.1007/s10208-010-9077-x} {\bibfield  {journal} {\bibinfo  {journal} {Foundations of Computational Mathematics}\ }\textbf {\bibinfo {volume} {11}},\ \bibinfo {pages} {65} (\bibinfo {year} {2011})}\BibitemShut {NoStop}%
\bibitem [{\citenamefont {Bernardi}\ \emph {et~al.}(2018)\citenamefont {Bernardi}, \citenamefont {Carlini}, \citenamefont {Catalisano}, \citenamefont {Gimigliano},\ and\ \citenamefont {Oneto}}]{bernardi2018hitchhiker}%
  \BibitemOpen
  \bibfield  {author} {\bibinfo {author} {\bibfnamefont {A.}~\bibnamefont {Bernardi}}, \bibinfo {author} {\bibfnamefont {E.}~\bibnamefont {Carlini}}, \bibinfo {author} {\bibfnamefont {M.~V.}\ \bibnamefont {Catalisano}}, \bibinfo {author} {\bibfnamefont {A.}~\bibnamefont {Gimigliano}}, \ and\ \bibinfo {author} {\bibfnamefont {A.}~\bibnamefont {Oneto}},\ }\href {\doibase 10.3390/math6120314} {\bibfield  {journal} {\bibinfo  {journal} {Mathematics}\ }\textbf {\bibinfo {volume} {6}},\ \bibinfo {pages} {314} (\bibinfo {year} {2018})}\BibitemShut {NoStop}%
\bibitem [{\citenamefont {Kanev}(1999)}]{kanev1999chordal}%
  \BibitemOpen
  \bibfield  {author} {\bibinfo {author} {\bibfnamefont {V.}~\bibnamefont {Kanev}},\ }\href {\doibase 10.1007/BF02367252} {\bibfield  {journal} {\bibinfo  {journal} {Journal of Mathematical Sciences}\ }\textbf {\bibinfo {volume} {94}},\ \bibinfo {pages} {1114} (\bibinfo {year} {1999})}\BibitemShut {NoStop}%
\bibitem [{\citenamefont {Pankowski}\ \emph {et~al.}(2010)\citenamefont {Pankowski}, \citenamefont {Piani}, \citenamefont {Horodecki},\ and\ \citenamefont {Horodecki}}]{pankowski2010few}%
  \BibitemOpen
  \bibfield  {author} {\bibinfo {author} {\bibfnamefont {{\L}.}~\bibnamefont {Pankowski}}, \bibinfo {author} {\bibfnamefont {M.}~\bibnamefont {Piani}}, \bibinfo {author} {\bibfnamefont {M.}~\bibnamefont {Horodecki}}, \ and\ \bibinfo {author} {\bibfnamefont {P.}~\bibnamefont {Horodecki}},\ }\href {\doibase 10.1109/TIT.2010.2050810} {\bibfield  {journal} {\bibinfo  {journal} {IEEE Transactions on Information Theory}\ }\textbf {\bibinfo {volume} {56}},\ \bibinfo {pages} {4085} (\bibinfo {year} {2010})}\BibitemShut {NoStop}%
\end{thebibliography}


%%%%%%%%%%%%%%%%%%%%%%%%%%%%%%%%%%%%%%%%%%%%%%%%%%%%%%%
\end{document}